# Automated Spin-Assisted Layer-by-Layer Epitaxy Produces Highly Oriented Mixed-Linker MOF Thin Films

Eleonora Afanasenko,[1] Benedetta Marmiroli,[2] Behnaz Abbasgholi-NA,[3] Barbara Sartori[2,] Giovanni Birarda,[4] Chiaramaria Stani,[4] Matjaž Finšgar,[5] Peter E. Hartmann,[6] Mark Bieber,[7] Emma Walitsch,[7] Rolf Breinbauer,[7] Simone Dal Zilio,[3] Sumea Klokic[2*] and Heinz Amenitsch[2*]

1 CERIC-ERIC, Trieste, Italy

2 Institute of Inorganic Chemistry, University of Technology, Graz, Austria

3 Istituto Officina dei Materiali - CNR-IOM, Trieste, Italy

4 Elettra Sincrotrone Trieste, Trieste, Italy

5 Faculty of chemistry and chemical engineering, University of Maribor, Slovenia

6 Institute of Physical and Theoretical Chemistry, University of Graz, Austria

7 Institute of Organic Chemistry, University of Technology, Graz, Austria

Correspondence: sumea.klokic@tugraz.at; heinz.amenitsch@tugraz.at

Abstract

Control over crystallographic orientation in metal-organic framework (MOF) thin films is crucial for exploiting their anisotropic properties in sensing, catalysis, and separation. Achieving reproducible, highly oriented films remains challenging, especially for mixed-linker, pillared-layered frameworks. Here we present an automated, spin-assisted layer-by-layer liquid-phase epitaxy (LbL-LPE) strategy that enables rapid, ambient-condition fabrication of highly oriented, mixed-linker MOF thin films, demonstrated for $Zn_2BDC_2DABCO$ (BDC = terephthalate, DABCO = 1,4-diazabicyclo[2.2.2]octane). Correlative process monitoring by grazing-incidence wide-angle X-ray scattering (GIWAXS), grazing-angle infrared (GI-IR) and UV-Vis spectroscopy, contact-angle measurements, scanning electron microscopy (SEM), and time-of-flight secondary ion mass spectrometry (ToF-SIMS) ensure formation of uniform films with exceptional out-of-plane (001) orientation (degree of orientation >85%, Hermans parameter ~0.95) and excellent reproducibility. This strategy enables highly reproducible, high-throughput fabrication of orientation-controlled MOF thin films, providing a generalizable alternative to conventional LbL approaches. By enabling reproducible and entirely automated fabrication of highly anisotropic architectures, this work establishes a platform for integrating oriented MOFs into next-generation optoelectronic, sensing, and membrane devices where directional transport and ordered pore alignment are essential.

## 1. Introduction

The fabrication of oriented crystalline materials has become increasingly important for the development of films whose properties depend on alignment along a specific crystallographic axis.[1–3] A prominent example is given by metal-organic frameworks (MOFs), consisting of ordered networks of metal nodes and organic linkers, which offer exceptional compositional tunability, high surface area and porosity, rendering them attractive for gas storage, sensing, and separation.[4–7] For many of these applications, exact control over crystallographic orientation is essential because many MOF properties, including gas permeability,[8–10] optical properties,[11,12] or photo-induced structural dynamics,[13,14] exhibit directional dependence arising from lattice anisotropy. For instance, crystallographic orientation in MOF films enables optical birefringence and anisotropic fluorescence,[11,15] promotes enhanced directional gas permeation in membranes with preferential pore-channel alignment,[16] and governs structural dynamics in stimuli-responsive frameworks.[17,18] These examples highlight that crystallographic alignment can strongly enhance or suppress functional performance, making oriented MOF films critical for maximizing directional responses and enabling reliable integration into devices such as sensors, switches, and electronic components, where uniformity and reproducibility are required. [19–23]

To date, several strategies have been developed to grow oriented MOF films , including liquid-phase epitaxy (LPE),[24,25] layer-by-layer assembly (LbL),[26] substrate-seeded heteroepitaxy (SSH),[11,27,28] vapor-assisted conversion (VAC),[29] and various external-field-assisted approaches, as comprehensively reviewed elsewhere.[30–32] Each method offers distinct advantages and inherent limitations, with only some enabling controllable or even tunable crystallographic orientation. Within this context, layer-by-layer liquid-phase epitaxy (LbL-LPE) is particularly attractive due to its low chemical consumption and minimal solvent waste.[24,25,33] Among LbL-LPE approaches, the spin-assisted method introduced by Chernikova *et al*.[34] particularly stands out for reduced solvent use, minimized cross-contamination through direct deposition of reactants onto the substrate, precise thickness control *via* the number of deposition cycles, and shorter fabrication times compared to other LPE methods.

Despite these strengths, spin-assisted LbL-LPE has not yet been translated into a robust high-throughput workflow that ensures reproducible orientation control and allows real-time monitoring and adjustment of critical process parameters - capabilities that are crucial for transferability across different MOF chemistries.[27,35] So far, only a limited number of studies have explored spin-assisted LbL approaches for MOF film fabrication, primarily focusing on frameworks with a single ligand type, such as ZIF-8, HKUST-1, and $Cu_2(bdc)_2 \cdot xH_2O$, SURMOF-2-type[36] or bimetallic MOF films.[37] In contrast, mixed-ligand or pillar-layered frameworks, highly relevant in the design of structurally adaptive materials, remain largely unexplored.[6,8] In these systems, orientation control is particularly important for anisotropic pore contraction[11] and light-responsive behaviour,[6,9] when guest-induced deformations[7] can propagate differently along distinct crystallographic directions, ultimately affecting the macroscopic functionality of the film.[8]

Here, we address these challenges by developing an automated, spin-assisted LbL-LPE protocol with integrated monitoring of all critical processing steps, enabling the reproducible fabrication of homogeneous and highly oriented MOF films (**Figure 1**). The workflow incorporates contact-angle analysis to verify self-assembled monolayer (SAM) functionalization and UV-visible spectroscopy to accurately adjust reactant concentrations. The structural quality and orientation of the resulting films are assessed through a correlative post-synthetic characterization strategy combining grazing-incidence wide-angle X-ray scattering (GIWAXS), grazing-incidence reflectance (GI-IR) spectromicroscopy, scanning electron microscopy (SEM), and time of flight secondary-ion mass spectrometry (ToF-SIMS). As a proof-of-concept system, we employ the flexible pillar-layered framework $Zn_2BDC_2DABCO$ (BDC = terephthalate, DABCO = 1,4-diazabicyclo[2.2.2]octane), chosen for its structural flexibility and previously demonstrated exceptional photoresponsive behavior and efficient

$CO_2$ uptake in SSH-grown oriented films as reported by some of the authors present herein.[4,13,38]

In the automated spin-assisted LbL-LPE process, individual deposition and rinsing steps are completed within seconds, enabling rapid film growth under ambient conditions with low reagent consumption. A modular software package enables automated execution and logging of each fabrication step, ensuring precise process control and facilitating implementation. By systematically varying the metal (Zn) to linkers (BDC:DABCO) ratios, we identify a regime that yields highly oriented films with the (001) reflection aligned parallel to the substrate. Deviations from this ratio lead to the emergence of partial (100) domains, as observed by GIWAXS and confirmed by SEM, highlighting the sensitivity of pillar-layered frameworks to reactant stoichiometry. Finally, the reproducible growth of homogeneous $Zn_2BDC_2DABCO$ films is achieved using both commercially available 16-mercaptohexadecanoic acid (MHDA) and the synthesized pyridine-terminated (4-(4-pyridyl)phenyl)methanethiol (PP1).[39]

The herein reported software-controlled spin-assisted LbL-LPE platform enables the reproducible fabrication of homogeneous, highly oriented MOF thin films, including complex pillar-layered and structurally adaptive frameworks. By combining precise crystallographic control with rapid, scalable deposition, this approach provides a reliable route to device-compatible MOF films and facilitates access to orientation-dependent properties for applications in gas separation, anisotropic sensing, optoelectronics, and stimuli-responsive actuation.

## 2. Synthesis of the $Zn_2(BDC)_2DABCO$ *via* LBL-LPE spin-coating

### 2.1 LPE spin-coating protocol

The automated LPE spin-coating protocol comprises three sequential steps, schematically outlined in **Figure 1**, each followed by a quality control check to ensure reproducible $Zn_2BDC_2DABCO$ thin film growth with controlled orientation, thickness, and uniformity.

The deposition process was carried out on gold substrates, pre-functionalized with self-assembled monolayers (SAMs), which provide the initial surface for MOF film growth. The spin-coating step is fully automated using a motorized arm equipped with three PTFE tubes, each connected to a separate syringe containing one of the working solutions: zinc acetate (M), the ethanolic mixed-linker BDC-DABCO solution (L), and ethanol as the washing solvent (S) (**Figure S1 a**). The syringes are mounted on syringe pumps, which deliver the working solutions dropwise onto the substrate through the tubing (**Figure S1 b**).

A single deposition cycle follows the sequence S-M-S-L-S and defines one growth cycle (**Figure S1 c**). $Zn_2BDC_2DABCO$ thin films were prepared using either 60 or 120 deposition cycles, with the number of repetitions and solution volumes fully programmable through LabVIEW-controlled automation (National Instruments, USA). The tube-to-substrate distance was optimized to achieve uniform coverage, and the resulting films were dried under a gentle nitrogen flow immediately following the last deposition cycle. This fully programmable setup allows flexible tuning of all critical parameters to achieve reproducible, high-quality MOF films. The specific optimization steps are discussed in detail in the following section.

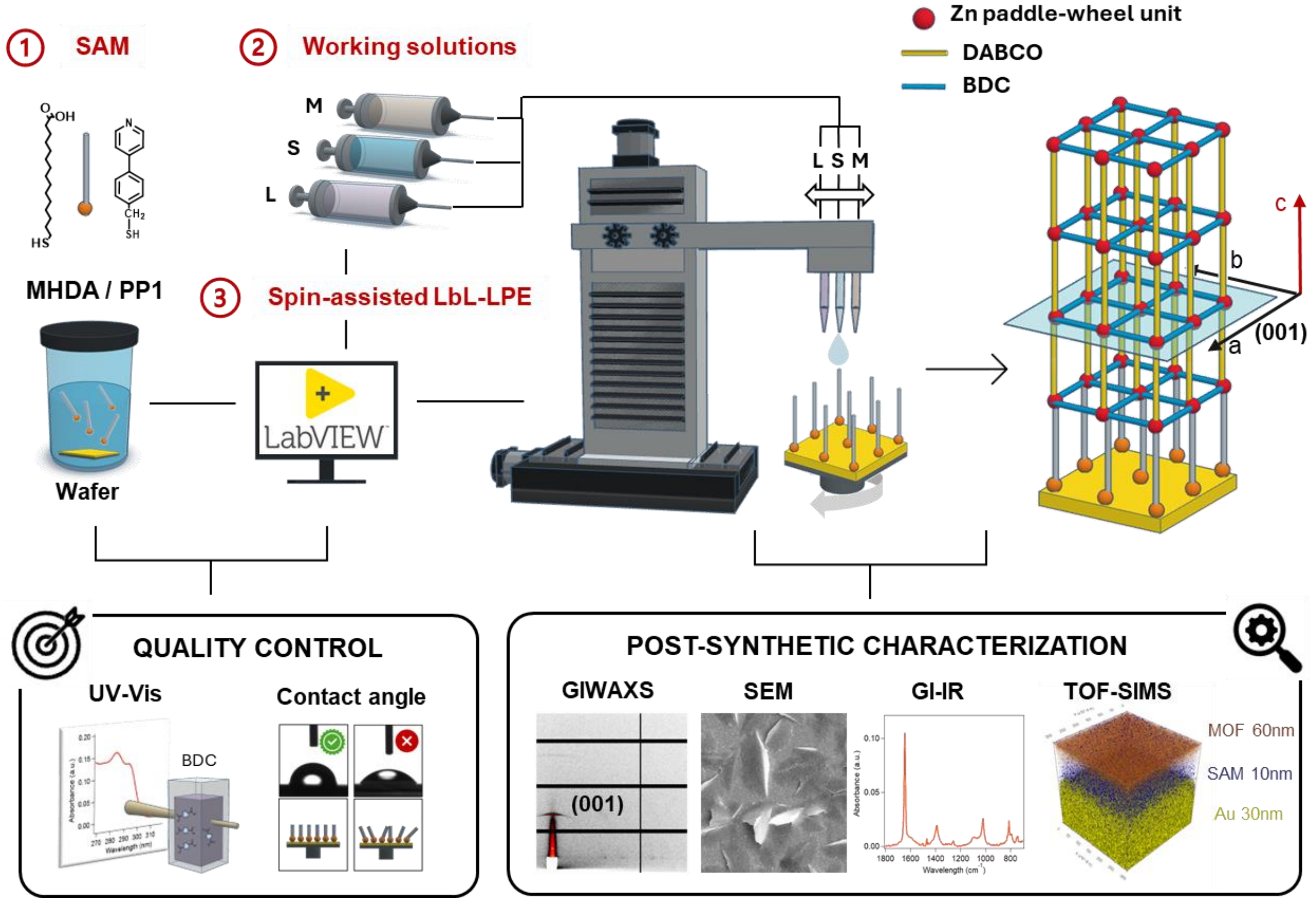

**Figure 1. Schematic representation of the spin-assisted LbL-LPE protocol.** The workflow comprises three key processing steps, each paired with quality-control or (post-synthetic) characterization: (1) gold wafers functionalization with MHDA or PP1 SAMs, with surface quality verified by contact angle measurements; (2) Preparation of precursor solutions, with concentrations confirmed by UV-Vis spectroscopy; (3) Fully automated layer-by-layer deposition controlled via LabVIEW, followed by structural characterization of the resulting thin films. The illustration represents the structure viewed along the (001) direction.

## 3. Results and discussion

### 3.1. Optimization of the LbL-LPE workflow

To ensure reproducibility and efficacy of the automated fabrication protocol, all parameters were carefully optimized and strict monitoring of critical steps, defined above as the quality control, was implemented after each process step (**Figure 1**, step 1-3). These key steps include firstly, the functionalization of gold wafers with MHDA or PP1 SAMs, with surface quality verified by contact angle measurements following well-established literature procedures.[6] Secondly, preparation of precursor solutions with concentrations confirmed by UV-Vis spectroscopy and, thirdly, the fully automated layer-by-layer deposition to grow oriented $Zn_2BDC_2DABCO$ films that are subsequently characterized structurally.

During the first step in the fabrication process (**Figure 1**, step 1), the gold wafers were functionalized with MHDA or PP1 SAMs, which serve as the anchoring layer for subsequent MOF growth. Here, only freshly SAM-functionalized substrates less than 48 hours old were employed for the LPE spin-coating process to ensure reproducible MOF thin film growth. The quality of the SAM was first confirmed by contact angle measurements (**Figure S2 a**), serving as the initial process quality control step. Generally, both SAM-forming molecules are known to be chemically stable across a broad range of experimental conditions and are widely used as anchoring layers in surface-mounted MOFs.[40–42] However, prolonged storage of functionalized wafers is known to induce surface disorder, which can adversely affect subsequent MOF growth.[43,44] Well-ordered MHDA layers exhibit pronounced hydrophilicity due to their terminal –COOH groups, as reflected by the measured contact angles ($\alpha$) of approximately $\alpha = 30°$, while defective or disordered SAMs display increased hydrophobicity with contact angles around $\alpha = 50°$ (**Figure S2 a**). A similar trend was observed for PP1 SAMs with $\alpha = 35°$ for well-ordered layers, and after loss of order values increase to $\alpha = 53°$, reflecting the combination of a hydrophilic pyridine moiety and a hydrophobic aromatic backbone. According to previous reports,[43–47] SAM degradation is attributed to UV-induced oxidation of the sulphur head groups and can be further promoted by both poor gold substrate quality or suboptimal functionalization procedures. Thus, for the process herein, restricting the use of substrates to those less than 48 hours old minimizes surface disorder and ensures reproducible crystallographic orientation and related film properties (*vide infra*).

During the second step in the fabrication process (**Figure 1**, step 2), focused on the preparation of the working solutions of the linkers ($H_2BDC$, DABCO) and metal ($Zn(ac)_2$), the ethanolic $H_2BDC$ solution is first analyzed by UV-Vis spectroscopy to verify its concentration. DABCO is added only afterward to the $H_2BDC$ solution, as the overlapping absorbance bands of both linkers preclude reliable differentiation in a single solution (**Figure S2 b**). DABCO in the separate solution was not detected, as its characteristic absorption band lies in the 190 nm region, close to the lower detection limit of the instrument.

Working solutions were typically used for no more than three days to avoid ligand crystallization or the formation of $Zn(OH)_2$.[48,49] It is important to note that one of the key factors influencing the quality of films produced by the LPE spin-coating process is the careful optimization of both the linker-to-linker and metal-to-linker ratios, which will be discussed in detail in Section 3.2. For the conditions used here, the nominal concentrations yielding reproducibly oriented and uniform MOF films were determined with Zn:BDC:DABCO = 0.1 mM : 0.1 mM : 0.3 mM.

Only after successful validation of step 1 (SAM functionalization) and step 2 (working solution preparation), the fully automated deposition process is carried out (**Figure 1**, step 3). During this step, 33 μL of each solution, determined empirically, are dispensed sequentially onto the functionalized substrate at a flow rate of 0.2 μL $min^{-1}$ while the substrate rotated continuously at 550 rpm. One layer is deposited within 40 s. The tube-to-substrate distance was fixed at 10 mm to control droplet impact and prevent coverage issues arising from the surface tension effect as visually inspected during process optimization: higher spinning speeds (800–1000 rpm) caused premature drying and prevented film growth, while lower speeds (450–500 rpm) led to unstable rotation and uneven coverage. Unlike conventional spin-assisted approaches involving intermittent drying, the optimized rotation speed of 550 rpm maintains a persistent liquid layer on the substrate, enabling continuous in-solution MOF self-assembly under precise stoichiometric control (see Chapter 3.2). Between each solution dispensing step, the sample is washed with ethanol to remove excess reagents and maintain uniform deposition. At completion of the deposition cycles, the film is removed from the spin-coater and dried under a gentle nitrogen stream. Leaving the wafer to dry under continuous spinning after completion of the deposition step results in a loss of crystalline orientation, leading to more isotropic GIWAXS patterns (**Figure S2 c**). In contrast, grazing-incidence infrared (GI-IR) spectromicroscopy confirmed that the chemical structure is preserved, as no differences are observed in the GI-IR spectra (**Figure S2 d**).

Following the optimized automated deposition conditions described above, the homogeneity and chemical composition of $Zn_2BDC_2DABCO$ films were assessed using GI-IR, which provides spatially resolved analysis across large substrate areas. To this aim, the entire wafer was raster-scanned, collecting individual GI-IR spectra distributed across the substrate. At a molar ratio of BDC:DABCO = 1:1:3, complete wafer coverage was achieved at the selected spinning speed without any edge inhomogeneities as commonly seen in spin-coated films (**Figure S3 a**). Also, other molar ratios were assessed by measuring dedicated areas in the center and on the sides (**Figure S3 b**) showing good homogeneity at 60 cycles, which was further verified by measuring GI-IR spectra of independently synthesized samples. The GIWAXS data further confirm that the grown MOF is homogeneous over the entire sample (**Figure S3 c**). The reproducibility of the automated LbL-LPE process was verified by quantifying crystallite orientation *via* the degree of orientation and Hermans orientation parameter, described in more detail in Section 3.3. Additional post-synthetic characterization, including the impact of variations in reagent molar ratios on morphology and crystallinity, is presented in the following section.

### 3.2. Structural and morphological characterization of LbL-LPE spin-coated films

Achieving reproducible, highly oriented MOF thin films under ambient conditions requires precise control over reactant stoichiometry and deposition parameters. This is because, the quality and orientation of MOF thin films critically depend on the relative amounts of linker and metal precursors, as dictated by fundamental principles of reactant stoichiometry and the roles of limiting and excess reagents in MOF assembly.[50–53] To identify the conditions that yield uniform, preferentially oriented films under the process conditions described in the Chapter 3.1, we systematically varied the BDC:DABCO molar ratio (1:1, 1:2, 1:3, 1:4, 1:5) and the number of deposition cycles (60 or 120). The resulting films were characterized by SEM and GIWAXS to assess morphology and crystallographic orientation, while their chemical composition was analyzed using GI-IR spectromicroscopy (**Figure S4**). GI-IR peak assignments (**Table S3**) were guided by DFT-calculated spectra, with computational details provided in the Supplementary Information (Chapter 5, **Figures S5–S11**, **Tables S1-S3)**.

We further examined how the linker ratio influences framework formation and crystallographic orientation in $Zn_2BDC_2DABCO$ thin films. GIWAXS measurements reveal a pronounced dependence of the crystallographic orientation on the BDC:DABCO ratio (**Figure 2a**). At a ratio of 1:1, MOF growth is suppressed on PP1-functionalized substrates, as evidenced by weak and diffuse reflections, while films grown on MHDA-functionalized substrates show no detectable diffraction. The GI-IR spectroscopy confirms the limited framework formation, with a peak at 1585 $cm^{-1}$ indicative of Zn-BDC coordination, but the absence of the C–N–C bending vibration at 1469 $cm^{-1}$ indicates insufficient DABCO incorporation (**Figure 2b**).

Increasing the relative DABCO content enhances framework formation, reflected by a sharp increase in the intensity of the (001) reflection. Films grown after 60 deposition cycles at ratios above 1:1 display a strong diffraction peak of the (001) lattice plane, indicating preferential out-of-plane alignment of the MOF lattice (**Figure S12 a**, for out-of-plane and in-plane directions). This alignment occurs independently of SAM functionalization, demonstrating that stoichiometric control dominates over surface chemistry in determining lattice orientation along the surface normal (**Figure 2a**). Comparison with the powder diffraction pattern of the bulk structure confirms that the film system adopts the tetragonal *P4/mmm* structure (**Figure S12 b**).[54] GI-IR spectromicroscopy further confirms coordination of the BDC linker, showing characteristic asymmetric and symmetric carboxylate stretching vibrations at 1624 $cm^{-1}$ and 1402 $cm^{-1}$, respectively (**Figure 2b**).

Films grown at BDC:DABCO ratios of 1:2 and 1:3 exhibit the highest out-of-plane orientation and uniformity, with the 1:3 ratio approaching a nearly single-crystal-like alignment, as evidenced by diffraction intensity focused into discrete spots (**Figure 2a**). When the DABCO content is increased beyond 1:3 (ratios 1:4 and 1:5), the azimuthal intensity distribution of the (001) reflection broadens over the azimuthal angle ($\chi$), reflecting reduced lattice alignment. The overall intensity of the (001) reflection decreases, indicating less well-ordered films. This is confirmed by GI-IR measurements, where spectra of the 1:5 ratio show a peak at 1294 $cm^{-1}$ corresponding to unreacted DABCO. For all films with ratios above 1:1, the dominant out-of-plane (001) reflection indicates preferential alignment of the c-axis perpendicular to the substrate. Notably, no evidence is observed for a well-defined in-plane registry, such as orthogonal alignment of (h00) and (0k0) planes across the surface, indicating that the films are only uniaxially along the out-of-plane direction (**Figure S12 c-e**).

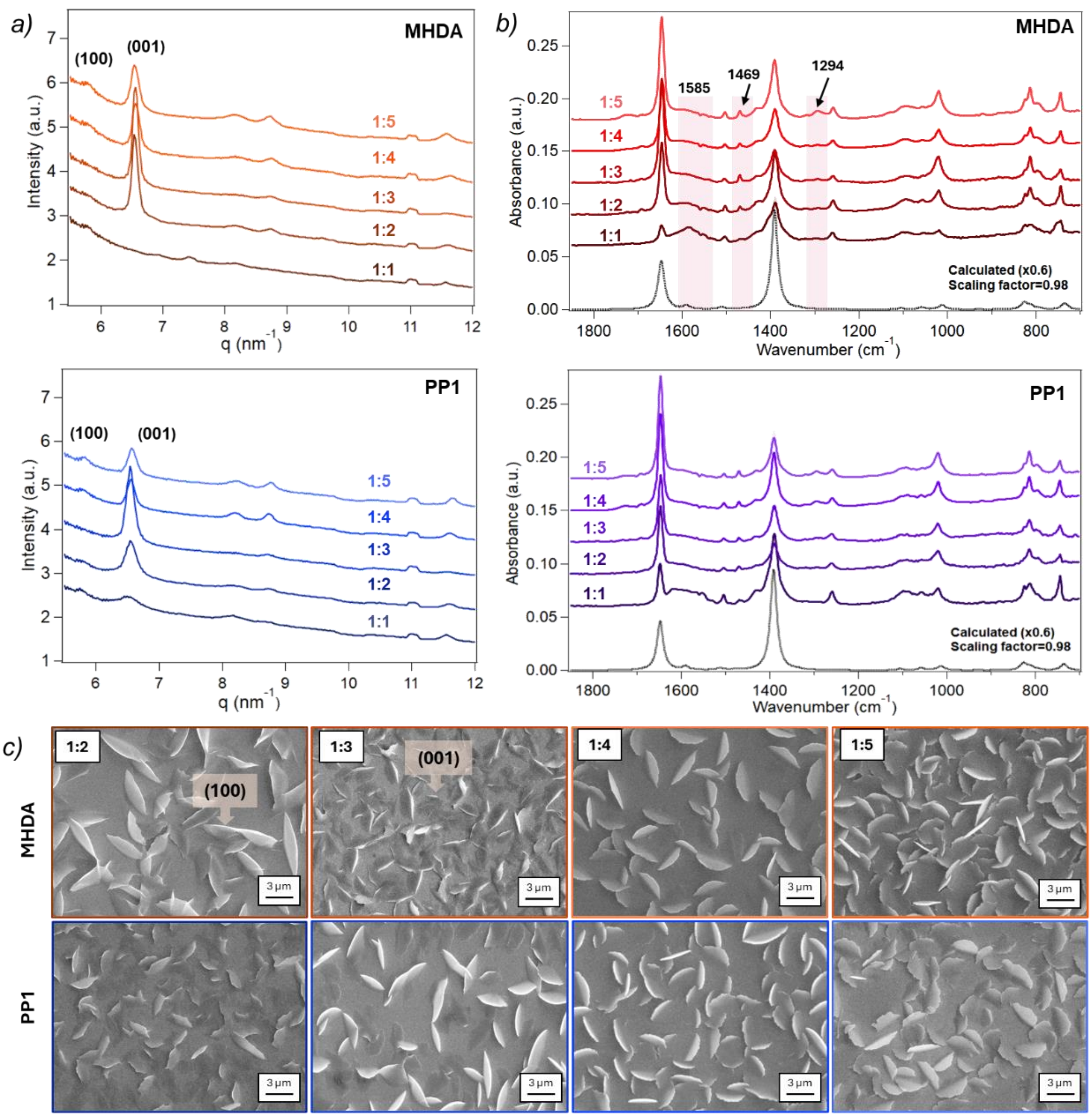

**Figure 2.** Characterization of $Zn_2BDC_2DABCO$ thin films after 60 cycles grown on MHDA and PP1 functionalized substrates as a function of BDC:DABCO molar ratio (1:1, 1:2, 1:3, 1:4, 1:5) a) Radially integrated GIWAXS patterns; b) GI-IR-spectra of the films. The calculated spectrum is obtained from the V-exp. model structure. The band at 1294 $cm^{-1}$ corresponds to unreacted DABCO, while the band at 1469 $cm^{-1}$ indicates coordinated DABCO. The band at 1585 $cm^{-1}$ confirms coordination of Zn to BDC ligands. c) SEM images of MOF films deposited on MHDA (top) and PP1 (bottom), for different BDC:DABCO ratios, indicating (001) aligned crystalline plates and (100) ones perpendicular to the surface.

To examine the morphology of the $Zn_2BDC_2DABCO$ films and to correlate crystallite arrangement with the observed out-of-plane orientation SEM imaging was performed (**Figure 2c**). At BDC:DABCO ratios of 1:2 and 1:3, the films comprise densely packed, plate-like crystallites aligned parallel to the substrate, consistent with the dominant (001) out-of-plane orientation observed in GIWAXS (*vide supra,* **Figure 2a**). In contrast, films with higher DABCO content (1:4 and 1:5) display crystallites of similar shape and thickness but with a broader distribution of tilt angles relative to the substrate, giving rise to partial (100) crystallographic domains (**Figure 2a**). This mixed-orientational growth behavior is similar to that previously reported in $Cu_2(F_4BDC)_2(DABCO)$,[40] where plate-like (001)-oriented structures were achieved by tuning the synthesis temperature. In that system, twinning within the paddle-wheel nodes was proposed as a mechanism for deviations from perfect (001) orientation.[40] In the present $Zn_2BDC_2DABCO$ films, a similar mechanism can rationalize the observations: at low DABCO content (1:2), insufficient coordination allows some $Zn_2$ paddle-wheel units in the upper layers to rotate around the σ-bonds of the BDC linker. These rotated units act as nucleation sites, introducing distortions into initially (001)-oriented crystallites, leading to twinning. At the optimal ratio of 1:3, this behavior is largely suppressed, with more than 85% of crystallites aligning along the (001) direction (*vide infra*, Section 3.3). Conversely, excessive DABCO (1:4–1:5) competes with $Zn^{2+}$ binding, which disrupts the formation of Zn–BDC 2D sheets and promotes more isotropic crystal orientations, consistent with both SEM and GIWAXS observations (**Figures 2a, c**). This results are also consistent with previous observations in $Cu_2(BDC)_2(DABCO)$ which is isostructural to the model system $Zn_2(BDC)_2(DABCO)$.[55]

Increasing the number of deposition cycles to 120 further affects crystallographic orientation. A weak peak at $q$ = 5.8 nm$^{-1}$ emerges, indicative of (100)-oriented domains (**Figure 3a**), whereas films grown with 60 cycles show no additional peaks (**Figure 2a**). This trend occurs irrespective of the BDC:DABCO ratio and likely arises from the cumulative effect of sequential layer growth on already twinned or distorted layers (**Figure 3b**). No significant differences are observed in the GI-IR spectra of samples prepared with 120 cycles compared to those with 60 cycles, apart from an overall increase in peak intensity (**Figure 3 c**), indicating that the chemical structure remains unchanged. These observations highlight that both stoichiometric balance and the number of deposition cycles are key parameters for achieving highly oriented, uniform MOF films with minimal defects.

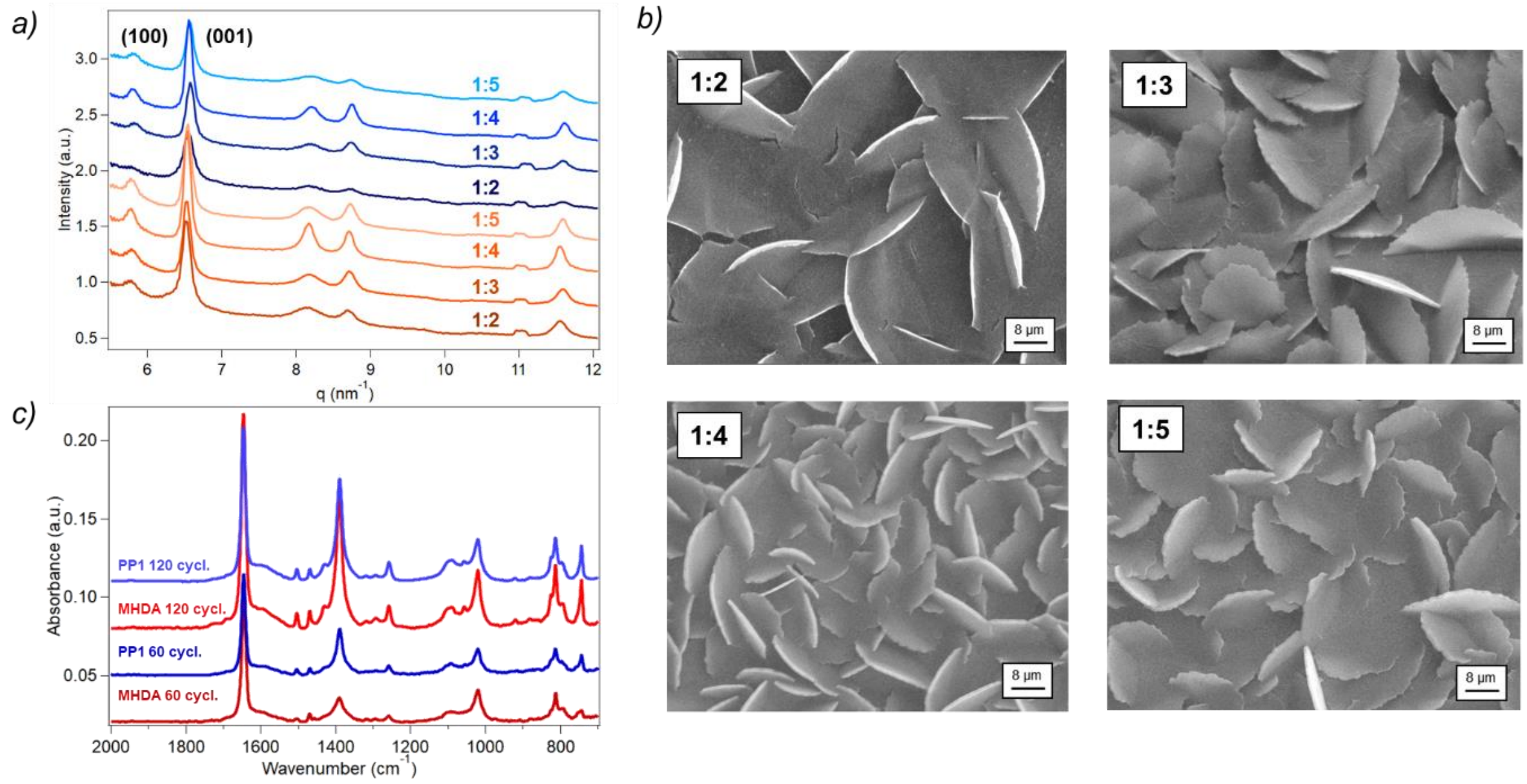

**Figure 3.** Characterization of $Zn_2BDC_2DABCO$ thin films after 120 cycles. a) Radially integrated GIWAXS patterns on MHDA (yellow) and PP1 (blue) functionalized substrates; b) SEM images of the films with different molar ratios; c) GI-IR-spectra of the BDC : DABCO = 1:3 ratio.

Cross-sectional SEM images prepared by FIB-SEM (**Figure S13 a, b**) show that both (001)- and (100)-oriented crystallites have comparable thicknesses (~64 nm) and grow directly on top of the SAM layer. Early-stage (001)-oriented crystals serve as seeding nodes for the subsequent growth of tilted and (100)-oriented crystallites. EDS analysis further indicates that the regions between crystallites contain only minor metal (**Figure S13 b**), suggesting that film formation proceeds *via* isolated nucleation sites rather than continuous layer growth, consistent with a Volmer-Weber growth mechanism.[15,55,56] Additional SEM images of samples with an equimolar BDC:DABCO ratio confirm their amorphous structure (**Figure S13 c**).

To further probe the film composition, ToF-SIMS analysis was performed. Three-dimensional reconstructions reveal $NH^-$, $C_6H_5^-$, $Au_3^-$, and $S^-$ signals, corresponding to nitrogen-containing fragments, aromatic moieties, the gold substrate, and sulfur species from the SAM, respectively (**Figure 4 a, b**). Depth profiling using a gas cluster ion beam (GCIB) shows a largely uniform elemental distribution for films prepared at BDC:DABCO ratios of 1:3 and 1:5 (MHDA-functionalized), with an increased $NH^-$ signal in the 1:5 sample (**Figure 4c**) reflecting excess DABCO, in agreement with GI-IR spectroscopy (**Figure 2b**). Sulphur-containing species are localized at the gold/MOF interface, confirming the presence of the SAM, while Zn-related signals co-localize with aromatic ($C_6H_5^-$) fragments, indicating homogeneous incorporation of Zn within the MOF framework. Two-dimensional images of the $ZnOH^-$ signal is homogeneously distributed across the scanned area, reflecting the presence of Zn–O(H) species within the film (**Figure 4d**). The formation of $Zn(OH)_2$ is unlikely, as its characteristic signals would produce clear features in the GI-IR spectrum of the fabricated films, which were not observed (at 3300 $cm^{-1}$). Therefore, the $ZnOH^-$ signal is more likely generated from ZnO *via* hydrogen attachment during $Bi_3^+$ bombardment or post-ionization processes and should be considered a marker of Zn–O(H) species rather than a unique identifier of a single phase. The $ZnOH^-$ coverage of the scanned area was calculated to be 50.6% for the sample with the molar ratio BDC:DABCO = 1:3 and 19.8% for the one with the ratio BDC:DABCO = 1:5. These results are consistent with the EDS analysis, which indicates that the crystals are spatially separated across the SEM-imaged surface.

Film thickness was determined by 3D profilometry of sputtered craters (**Figure S14**), yielding a total thickness of 130 ± 15 nm for representative samples. This includes ~50 nm of Au, ~10 nm of SAM, and ~70 nm of MOF, in good agreement with the crystallite dimensions observed by SEM.

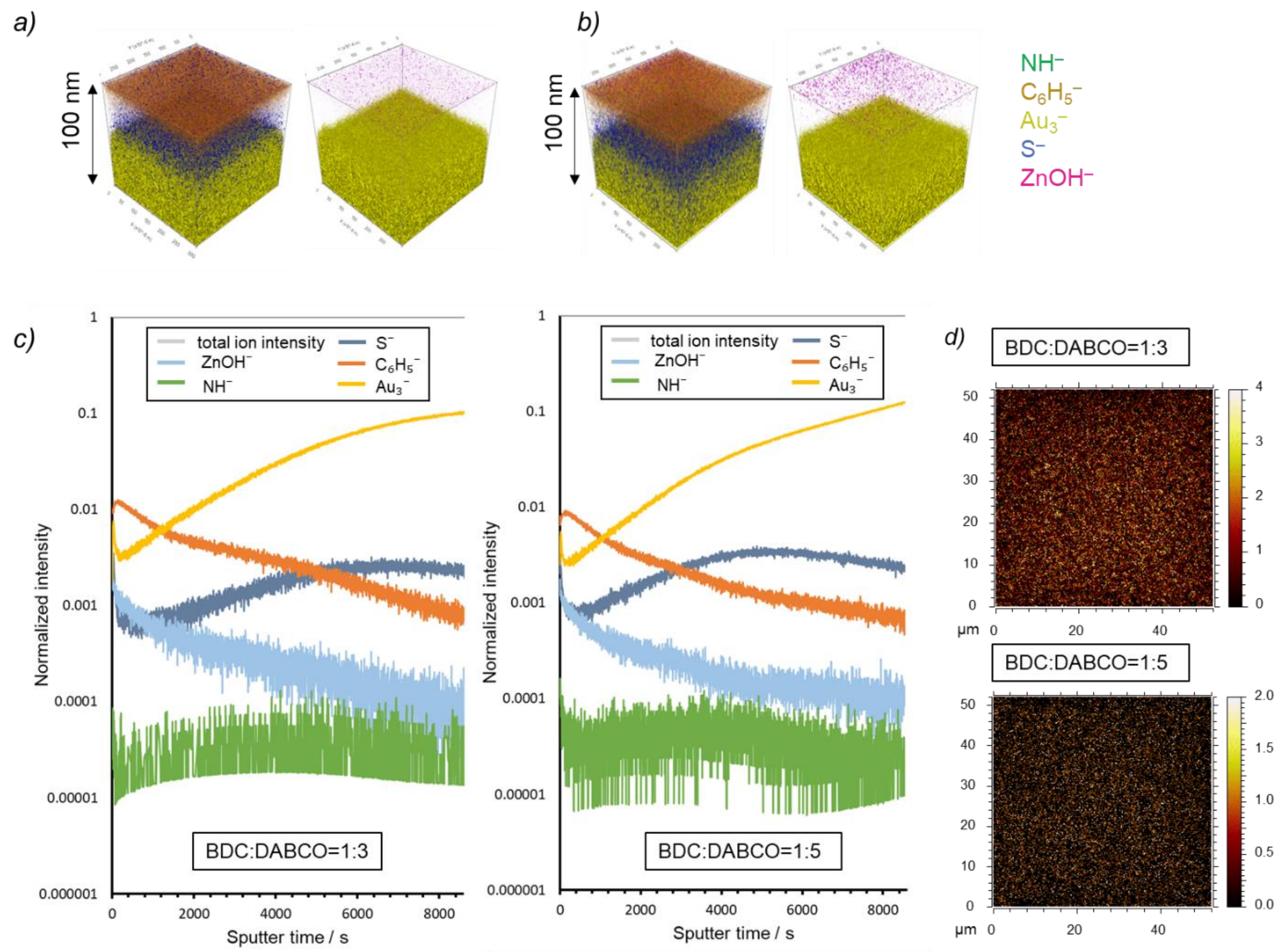

**Figure 4.** Characterization of $Zn_2BDC_2DABCO$ thin films by ToF-SIMS. Three-dimensional ToF-SIMS reconstructions of samples with the molar ratios Zn:BDC:DABCO = 1:1:3 (a) and 1:1:5 (b) on the MHDA self-assembled monolayer (60 cycles), confirming the thickness measured from SEM images. c) GCIB depth profiles showing the depth distribution of $NH^-$, $C_6H_5^-$, $Au_3^-$, $S^-$, and $ZnOH^-$ signals. (d) Two-dimensional ToF-SIMS images showing the distribution of the $ZnOH^-$ signal on the MOF film surface (topmost position).

Building on these observations, the influence of the metal-to-ligand ratio on film orientation was systematically investigated for samples grown for 60 deposition cycles on MHDA-functionalized substrates (**Figure 5**). Increasing the metal concentration at a BDC:DABCO ratio of 1:1 did not result in crystal formation, consistent with the previously identified limitation of insufficient linker content (**Figure 5a**). At the optimal ligand ratio of 1:3, reducing the zinc acetate concentration to 0.05 mM yielded well-oriented (001) films, albeit with incomplete substrate coverage, particularly at the sample edges, due to insufficient material deposition, as confirmed by SEM and GI-IR spectroscopy (**Figure 5b, c**). Increasing the metal concentration to 0.2 mM resulted in uniform, highly oriented films, whereas further increases beyond 0.3 mM led to a loss of preferential orientation, as evidenced by the GIWAXS patterns (**Figure 5a**).

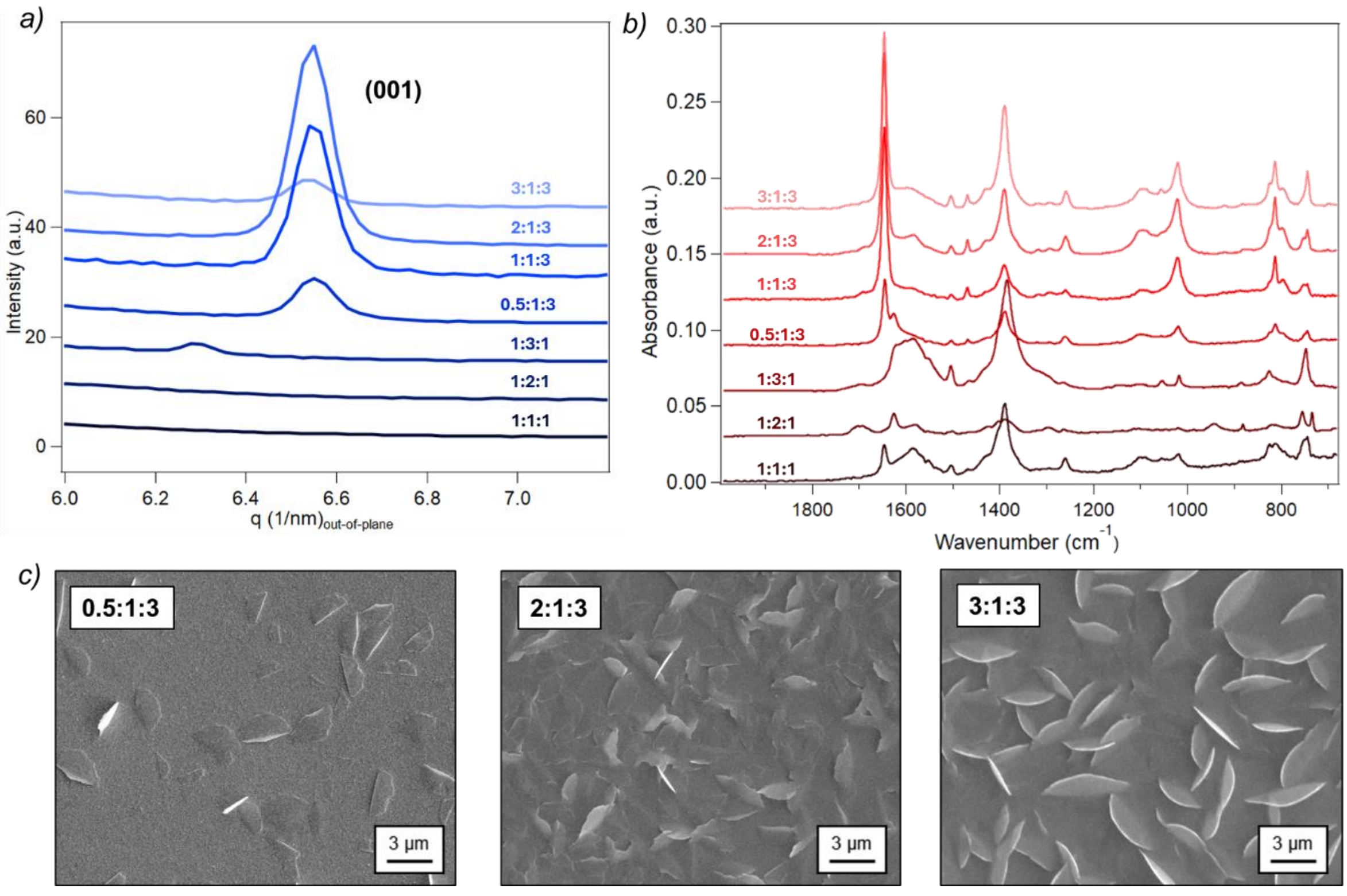

**Figure 5.** Characterization of samples prepared with 120 deposition cycles at varying Zn-to-ligand ratios on an MHDA self-assembled monolayer: (a) GIWAXS, out-of-plane cut to compare (001) reflection, (b) GI-IR, and (c) SEM data of the samples prepared by the variation of the Zn-BDC-DABCO molar ratios.

### 3.3. Quantitative analysis of film quality and reproducibility evaluation

To quantify lattice alignment and assess reproducibility, the degree of orientation (DO) and Herman's orientation parameter (HOP) were evaluated from GIWAXS data. A detailed description of the analysis procedure is provided in the Supplementary Information (Chapter 9, **Figure S15**).[57,58] While both parameters describe crystallite orientation, they reflect complementary aspects: DO quantifies the fraction of aligned crystallites within the film with higher values indicating more homogeneous, well-aligned films, whereas lower values point to increased misalignment or heterogeneity. HOP provides a quantitative measure of the degree of preferential orientation based on the azimuthal intensity distribution. HOP values range from 1 (perfectly aligned along the reference axis) to 0 (random orientation) and -0.5 (perfect perpendicular alignment). Both orientation parameters provide a clear measure of how the MOF lattice aligns with respect to the substrate and, because GIWAXS probes a relatively large area, they also reflect the overall homogeneity of the films. When combined with peak-width analysis (full width at half maximum, FWHM), they provide complementary information on film quality and enable evaluation of reproducibility across samples, capturing both macroscopic texture and nanoscale crystallite arrangement. Unlike one-dimensional out-of-plane GIWAXS scans or SEM micrographs, this approach quantitatively distinguishes oriented from isotropically distributed crystallite populations. Despite appearing similar in conventional analyses, MHDA- and PP1-functionalized films are differentiated by DO and HOP, which reveal subtle structural differences and are discussed in detail in the following.

The extracted HOP and DO mean values for MHDA- and PP1-functionalized films are summarized in **Figure 6 a, b** and Table S4. For films prepared at a BDC:DABCO ratio of 1:3 on MHDA-functionalized substrates, excellent alignment is achieved with DO = 85 ± 10% (**Figure S16**) and HOP values approaching unity (0.93), accompanied by narrow FHWM values. This functionalization and reactant ratio marks the optimum conditions and indicates that most crystallites are oriented with their (001) planes parallel to the substrate. At lower ratios (1:2), both DO and HOP decrease and exhibit a larger spread amongst different samples, reflecting reduced alignment and poorer reproducibility. Increasing the DABCO content beyond the optimal ratio 1:4 and 1:5 results only in minor changes of the DO and HOP values, but a pronounced broadening of the azimuthal distribution, indicating a wider spread of crystallite orientations and reduced alignment sharpness.

Despite similar GIWAXS patterns and SEM morphologies, quantitative analysis reveals subtle differences between SAM functionalizations: Films grown on PP1-functionalized substrates at a ratio of 1:3 exhibit lower DO values (61.30 ± 23.70%) and larger variability compared to MHDA-functionalized films (88.64 ± 7.94%), indicating reduced reproducibility, while HOP values remain comparable (0.95). Increasing the number of deposition cycles to 120 for MHDA and PP1 films does not significantly affect the mean DO, whilst HOP decreases along with a clear increase in FWHM (**Figure 6b**), consistent with a higher fraction of tilted crystallites as previously inferred from SEM and GIWAXS analysis (*vide supra*, **Figure 3**). Based on the combined GIWAXS and spectroscopic data, three distinct alignment regimes can be identified for the herein fabricated $Zn_2BDC_2DABCO$ films *via* the spin-assisted automated LbL-LPE process, schematically summarized in **Figure 6d**.

For BDC-rich conditions, MOF growth is suppressed, where the limited structure formed being largely amorphous. Gradually increasing the DABCO concentration up to the optimal stoichiometry (BDC:DABCO = 1:3) produces highly oriented, uniform films dominated by (001)-oriented crystallites with narrow azimuthal distributions are obtained (DO > 85%, HOP ≈ 0.9–1.0). In contrast, excess DABCO leads to broader orientation distributions and the emergence of mixed (001)/ (100) domains, consistent with tilted crystallite growth observed by SEM. These regimes are reflected in the corresponding 2D GIWAXS patterns (**Figure 6c**), where sharp, localized reflections indicate high alignment, whereas azimuthally broadened intensity distributions signify increased disorder.

Overall, a BDC:DABCO ratio of 1:3 on MHDA-functionalized substrates with 60 deposition cycles yields the most favorable combination of high orientation, uniformity, and reproducibility. Notably, the herein achieved DO (~85%) compares favorably with previously reported epitaxial $Zn_2BDC_2DABCO$ films seeded either on $Cu(BDC)_2$ (DO ~ 45%)[11] or on $Cu_2BDC_2DABCO$ (~93%).[28] This demonstrates that the spin-assisted LbL-LPE approach produces highly oriented MOF films. The ability to achieve well-aligned films simply by adjusting the linker-to-linker and metal-to-linker ratios demonstrates that commercially available COOH-terminated thiols can be used directly for LbL-LPE growth of $Zn_2BDC_2DABCO$, eliminating the need for pyridine-terminated SAMs, which require a lengthy synthesis and extensive purification (Supplementary information, chapter 1.3).[37]

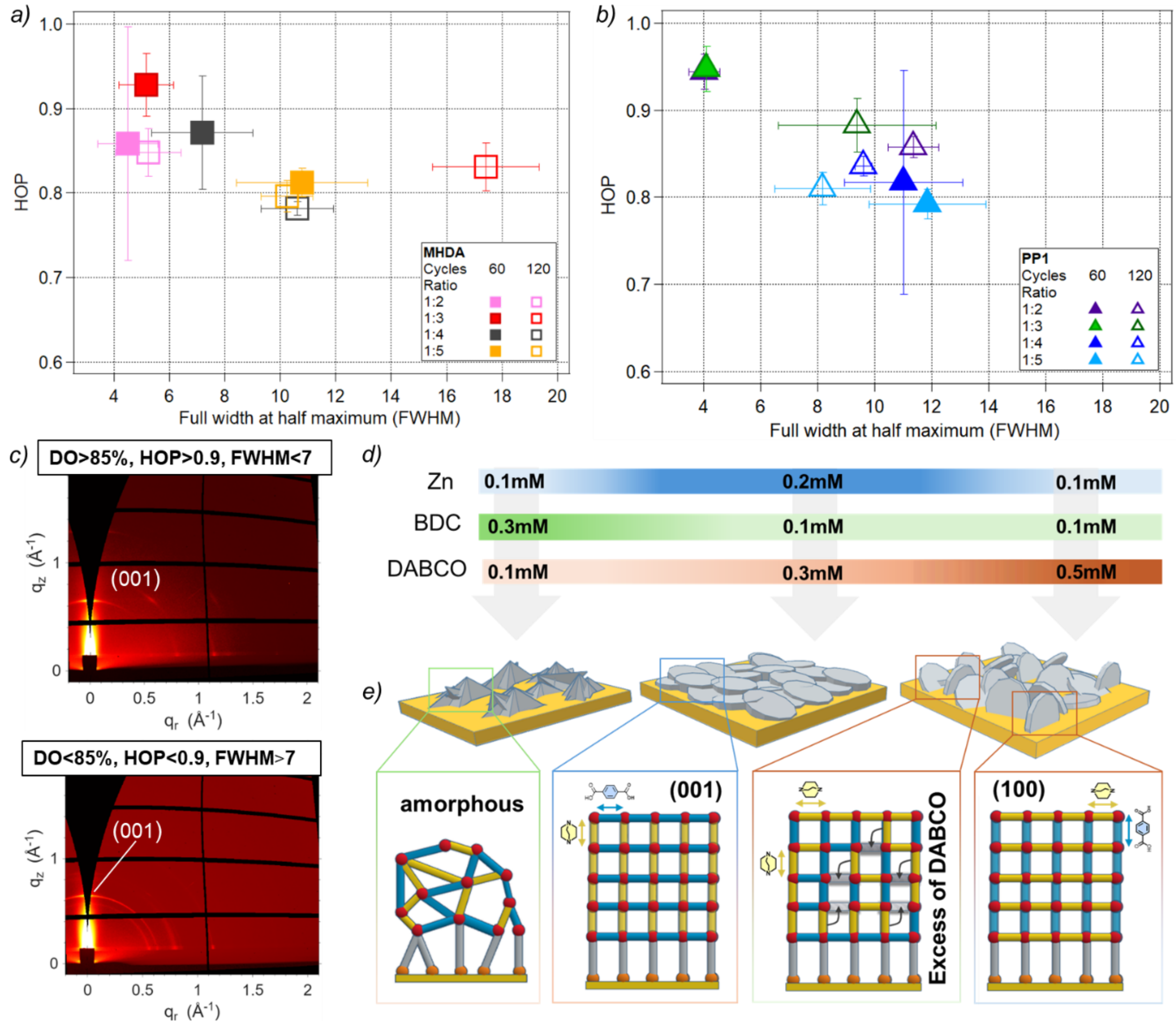

**Figure 6.** HOP versus full-width at half maximum (FWHM) of the (001) reflection, where filled markers indicate 60 cycles and unfilled markers 120 cycles for (a) MHDA- and (b) PP1 functionalized samples. The HOP and corresponding peak widths were averaged from triplicate samples (and from eight samples for the 1:3 ligand ratio at 60 cycles) under each synthetic condition. (c) Reshaped GIWAXS images showing highly oriented and isotropic scattering patterns. (d) Schematic representation of how changes in the molar ratio of reagents influence the crystalline orientation of LbL-LPE-synthesized films. (e) Illustration of different crystalline

orientations and the mechanism of isotropic growth arising from the excess of DABCO ligand in the MOF structures on the substrate.

**4. Conclusion**

This work demonstrates the fully automated fabrication of mixed-linker, pillared-layered MOF thin films *via* spin-assisted LbL-LPE under ambient conditions. Using $Zn_2BDC_2DABCO$ as a model system, we show that precise control over linker-to-linker and metal-to-linker ratios enables the reproducible growth of homogeneous and highly oriented films without elevated temperatures or toxic solvents, supporting sustainable and low-consumption film formation. The automated LbL-LPE workflow provides fine control over key processing parameters, including SAM quality, reactant concentrations, deposition volume, and rotational speed, thereby enabling reliable tuning of film orientation and morphology. Correlative characterization by GIWAXS, GI-IR, and SEM confirms exceptional alignment along the surface normal together with homogeneous substrate coverage. An optimal Zn:BDC:DABCO ratio of 1:1:3 yields uniformly covered films with strong out-of-plane (001) alignment (DO > 85%, HOP ~ 0.95).

Beyond this model system, the results demonstrate that spin-assisted LbL-LPE is applicable to the fabrication of oriented MOF thin films, including structurally complex mixed-linker architectures. The combination of automation, ambient processing, and low reagent consumption addresses key challenges in reproducibility and sustainable manufacturing of MOF thin films across laboratories. The high degree of crystallographic orientation is particularly relevant for orientation-dependent applications, such as membranes, sensors, and optoelectronic devices. Overall, this work establishes a strategy for the controlled fabrication of highly ordered MOF thin films, aiming to facilitate their integration into functional device architectures.

**Acknowledgements**

The authors acknowledge Ing. Andrea Radeticchio for technical support. The authors acknowledge the CERIC-ERIC consortium for support through the project for postdoctoral researchers from Ukraine affected by the war, as well as access to the Austrian SAXS beamline at Elettra Sincrotrone Trieste through proposals 20247158, 20247199, 20242148, and 20242144. The authors further acknowledge SISSI-Bio@Elettra for providing in-house beamtime. Financial support from the Slovenian Research Agency (P2-0118) is gratefully acknowledged. The project was co-financed by the Republic of Slovenia, the Ministry of Education, Science and Sport, and the European Union through the European Regional Development Fund. Part of this work was supported by the Interreg Italy–Slovenia Program (2021-2027, AllMICRO project). The authors also thank Gregor Kapun of the Nanocenter for valuable support with the cross-sectional FIB-SEM analyses.

**Data Availability Statement**

((include as appropriate, including link to repository))

**References**

(1) Khalil, I. E.; Fonseca, J.; Reithofer, M. R.; Eder, T.; Chin, J. M. Tackling Orientation of Metal-Organic Frameworks (MOFs): The Quest to Enhance MOF Performance. *Coordination Chemistry Reviews* **2023**, *481*, 215043. https://doi.org/10.1016/j.ccr.2023.215043.

(2) Gilroy, K. D.; Ruditskiy, A.; Peng, H.-C.; Qin, D.; Xia, Y. Bimetallic Nanocrystals: Syntheses, Properties, and Applications. *Chem. Rev.* **2016**, *116* (18), 10414–10472. https://doi.org/10.1021/acs.chemrev.6b00211.

(3) Brandner, L. A.; Marmiroli, B.; Linares-Moreau, M.; Barella, M.; Abbasgholi-NA, B.; Velásquez-Hernández, M. D. J.; Flint, K. L.; Dal Zilio, S.; Acuna, G. P.; Wolinski, H.;

Amenitsch, H.; Doonan, C. J.; Falcaro, P. Ordered Transfer from 3D-Oriented MOF Superstructures to Polymeric Films: Microfabrication, Enhanced Chemical Stability, and Anisotropic Fluorescent Patterns. *Advanced Materials* **2024**, *36* (46), 2404384. https://doi.org/10.1002/adma.202404384.

(4) Klokic, S.; Marmiroli, B.; Birarda, G.; Lackner, F.; Holzer, P.; Sartori, B.; Abbasgholi-NA, B.; Dal Zilio, S.; Kargl, R.; Stana Kleinschek, K.; Stani, C.; Vaccari, L.; Amenitsch, H. Flexible Metal-Organic Framework Films for Reversible Low-Pressure Carbon Capture and Release. *Nat Commun* **2025**, *16* (1), 7135. https://doi.org/10.1038/s41467-025-60027-6.

(5) ZareKarizi, F.; Joharian, M.; Morsali, A. Pillar-Layered MOFs: Functionality, Interpenetration, Flexibility and Applications. *J. Mater. Chem. A* **2018**, *6* (40), 19288–19329. https://doi.org/10.1039/C8TA03306D.

(6) Wannapaiboon, S.; Schneemann, A.; Hante, I.; Tu, M.; Epp, K.; Semrau, A. L.; Sternemann, C.; Paulus, M.; Baxter, S. J.; Kieslich, G.; Fischer, R. A. Control of Structural Flexibility of Layered-Pillared Metal-Organic Frameworks Anchored at Surfaces. *Nat Commun* **2019**, *10* (1), 346. https://doi.org/10.1038/s41467-018-08285-5.

(7) Zhang, J.-P.; Zhou, H.-L.; Zhou, D.-D.; Liao, P.-Q.; Chen, X.-M. Controlling Flexibility of Metal–Organic Frameworks. *National Science Review* **2018**, *5* (6), 907–919. https://doi.org/10.1093/nsr/nwx127.

(8) Senkovska, I.; Bon, V.; Abylgazina, L.; Mendt, M.; Berger, J.; Kieslich, G.; Petkov, P.; Luiz Fiorio, J.; Joswig, J.; Heine, T.; Schaper, L.; Bachetzky, C.; Schmid, R.; Fischer, R. A.; Pöppl, A.; Brunner, E.; Kaskel, S. Understanding MOF Flexibility: An Analysis Focused on Pillared Layer MOFs as a Model System. *Angew Chem Int Ed* **2023**, *62* (33), e202218076. https://doi.org/10.1002/anie.202218076.

(9) Jin, E.; Bon, V.; Das, S.; Wonanke, A. D. D.; Etter, M.; Karlsen, M. A.; De, A.; Bönisch, N.; Heine, T.; Kaskel, S. Engineering Photoswitching Dynamics in 3D Photochromic Metal–Organic Frameworks through a Metal–Organic Polyhedron Design. *J. Am. Chem. Soc.* **2025**, *147* (10), 8568–8577. https://doi.org/10.1021/jacs.4c17203.

(10) Baumgartner, B.; Mashita, R.; Fukatsu, A.; Okada, K.; Takahashi, M. Guest Alignment and Defect Formation during Pore Filling in Metal–Organic Framework Films. *Angew Chem Int Ed* **2022**, *61* (28), e202201725. https://doi.org/10.1002/anie.202201725.

(11) Falcaro, P.; Okada, K.; Hara, T.; Ikigaki, K.; Tokudome, Y.; Thornton, A. W.; Hill, A. J.; Williams, T.; Doonan, C.; Takahashi, M. Centimetre-Scale Micropore Alignment in Oriented Polycrystalline Metal–Organic Framework Films via Heteroepitaxial Growth. *Nature Mater* **2017**, *16* (3), 342–348. https://doi.org/10.1038/nmat4815.

(12) Vinogradov, A. V.; Milichko, V. A.; Zaake-Hertling, H.; Aleksovska, A.; Gruschinski, S.; Schmorl, S.; Kersting, B.; Zolnhofer, E. M.; Sutter, J.; Meyer, K.; Lönnecke, P.; Hey-Hawkins, E. Unique Anisotropic Optical Properties of a Highly Stable Metal–Organic Framework Based on Trinuclear Iron( III ) Secondary Building Units Linked by Tetracarboxylic Linkers with an Anthracene Core. *Dalton Trans.* **2016**, *45* (17), 7244–7249. https://doi.org/10.1039/C6DT00390G.

(13) Klokic, S.; Naumenko, D.; Marmiroli, B.; Carraro, F.; Linares-Moreau, M.; Zilio, S. D.; Birarda, G.; Kargl, R.; Falcaro, P.; Amenitsch, H. Unraveling the Timescale of the Structural Photo-Response within Oriented Metal–Organic Framework Films. *Chem. Sci.* **2022**, *13* (40), 11869–11877. https://doi.org/10.1039/D2SC02405E.

(14) Klokic, S.; Marmiroli, B.; Naumenko, D.; Birarda, G.; Dal Zilio, S.; Velásquez-Hernández, M. D. J.; Falcaro, P.; Vaccari, L.; Amenitsch, H. Orthogonal Stimulation of Structural Transformations in Photo-Responsive MOF Films through Linker Functionalization. *CrystEngComm* **2024**, *26* (17), 2228–2232. https://doi.org/10.1039/D4CE00221K.

(15) Ehrling, S.; Miura, H.; Senkovska, I.; Kaskel, S. From Macro- to Nanoscale: Finite Size Effects on Metal–Organic Framework Switchability. *Trends in Chemistry* **2021**, *3* (4), 291–304. https://doi.org/10.1016/j.trechm.2020.12.012.

(16) Zhai, M.; Moghadam, F.; Gosiamemang, T.; Heng, J. Y. Y.; Li, K. Facile Orientation Control of MOF-303 Hollow Fiber Membranes by a Dual-Source Seeding Method. *Nat Commun* **2024**, *15* (1), 10264. https://doi.org/10.1038/s41467-024-54730-z.

(17) Talin, A. A.; Centrone, A.; Ford, A. C.; Foster, M. E.; Stavila, V.; Haney, P.; Kinney, R. A.; Szalai, V.; El Gabaly, F.; Yoon, H. P.; Léonard, F.; Allendorf, M. D. Tunable Electrical Conductivity in Metal-Organic Framework Thin-Film Devices. *Science* **2014**, *343* (6166), 66–69. https://doi.org/10.1126/science.1246738.

(18) Okada, K.; Mori, K.; Fukatsu, A.; Takahashi, M. Oriented Growth of Semiconducting TCNQ@$Cu_3$ $(BTC)_2$ MOF on $Cu(OH)_2$ : Crystallographic Orientation and Pattern Formation toward Semiconducting Thin-Film Devices. *J. Mater. Chem. A* **2021**, *9* (35), 19613–19618. https://doi.org/10.1039/D1TA02968A.

(19) Xu, T.; Liu, Z.; Li, X.; Xu, T. The Evolution of Metal–Organic Framework Membranes: From Laboratory Innovation to Industrial Implementations. *Ind. Eng. Chem. Res.* **2025**, *64* (20), 9847–9866. https://doi.org/10.1021/acs.iecr.5c00781.

(20) Shrivastav, V.; Mansi; Gupta, B.; Dubey, P.; Deep, A.; Nogala, W.; Shrivastav, V.; Sundriyal, S. Recent Advances on Surface Mounted Metal-Organic Frameworks for Energy Storage and Conversion Applications: Trends, Challenges, and Opportunities. *Advances in Colloid and Interface Science* **2023**, *318*, 102967. https://doi.org/10.1016/j.cis.2023.102967.

(21) Haldar, R.; Heinke, L.; Wöll, C. Advanced Photoresponsive Materials Using the Metal–Organic Framework Approach. *Advanced Materials* **2020**, *32* (20), 1905227. https://doi.org/10.1002/adma.201905227.

(22) Kolodzeiski, E.; Amirjalayer, S. Dynamic Network of Intermolecular Interactions in Metal-Organic Frameworks Functionalized by Molecular Machines. *Science Advances* **2022**, *8* (26), eabn4426. https://doi.org/10.1126/sciadv.abn4426.

(23) Xiao, Y.-H.; Tian, Y.-B.; Gu, Z.-G.; Zhang, J. Surface-Coordinated Metal-Organic Framework Thin Films (SURMOFs): From Fabrication to Energy Applications. *EnergyChem* **2021**, *3* (6), 100065. https://doi.org/10.1016/j.enchem.2021.100065.

(24) Shekhah, O.; Wang, H.; Kowarik, S.; Schreiber, F.; Paulus, M.; Tolan, M.; Sternemann, C.; Evers, F.; Zacher, D.; Fischer, R. A.; Wöll, C. Step-by-Step Route for the Synthesis of Metal−Organic Frameworks. *J. Am. Chem. Soc.* **2007**, *129* (49), 15118–15119. https://doi.org/10.1021/ja076210u.

(25) Zhuang, J.-L.; Terfort, A.; Wöll, C. Formation of Oriented and Patterned Films of Metal–Organic Frameworks by Liquid Phase Epitaxy: A Review. *Coordination Chemistry Reviews* **2016**, *307*, 391–424. https://doi.org/10.1016/j.ccr.2015.09.013.

(26) Makiura, R.; Motoyama, S.; Umemura, Y.; Yamanaka, H.; Sakata, O.; Kitagawa, H. Surface Nano-Architecture of a Metal–Organic Framework. *Nature Mater* **2010**, *9* (7), 565–571. https://doi.org/10.1038/nmat2769.

(27) Linares-Moreau, M.; Brandner, L. A.; Kamencek, T.; Klokic, S.; Carraro, F.; Okada, K.; Takahashi, M.; Zojer, E.; Doonan, C. J.; Falcaro, P. Semi-Automatic Deposition of Oriented $Cu(OH)_2$ Nanobelts for the Heteroepitaxial Growth of Metal–Organic Framework Films. *Adv Materials Inter* **2021**, *8* (21), 2101039. https://doi.org/10.1002/admi.202101039.

(28) Zhao, T.; Taghizade, N.; Fischer, J. C.; Richards, B. S.; Howard, I. A. [001]-Oriented Heteroepitaxy for Fabricating Emissive Surface Mounted Metal–Organic Frameworks. *J. Mater. Chem. C* **2024**, *12* (15), 5496–5505. https://doi.org/10.1039/D4TC00018H.

(29) Romero-Ángel, M.; Rubio-Giménez, V.; Gómez-Oliveira, E. P.; Verstreken, M. F. K.; Smets, J.; Gándara-Loe, J.; M Padial, N.; Ameloot, R.; Tatay, S.; Martí-Gastaldo, C. Vapor-Assisted Conversion of Heterobimetallic Titanium–Organic Framework Thin Films. *Chem. Mater.* **2023**, *35* (24), 10394–10402. https://doi.org/10.1021/acs.chemmater.3c01389.

(30) Bétard, A.; Fischer, R. A. Metal–Organic Framework Thin Films: From Fundamentals to Applications. *Chem. Rev.* **2012**, *112* (2), 1055–1083. https://doi.org/10.1021/cr200167v.

(31) Sabzehmeidani, M. M.; Gafari, S.; jamali, S.; Kazemzad, M. Concepts, Fabrication and Applications of MOF Thin Films in Optoelectronics: A Review. *Applied Materials Today* **2024**, *38*, 102153. https://doi.org/10.1016/j.apmt.2024.102153.

(32) Shi, X.; Shan, Y.; Du, M.; Pang, H. Synthesis and Application of Metal-Organic Framework Films. *Coordination Chemistry Reviews* **2021**, *444*, 214060. https://doi.org/10.1016/j.ccr.2021.214060.

(33) Gu, Z.-G.; Zhang, J. Epitaxial Growth and Applications of Oriented Metal–Organic Framework Thin Films. *Coordination Chemistry Reviews* **2019**, *378*, 513–532. https://doi.org/10.1016/j.ccr.2017.09.028.

(34) Chernikova, V.; Shekhah, O.; Eddaoudi, M. Advanced Fabrication Method for the Preparation of MOF Thin Films: Liquid-Phase Epitaxy Approach Meets Spin Coating Method. *ACS Appl. Mater. Interfaces* **2016**, *8* (31), 20459–20464. https://doi.org/10.1021/acsami.6b04701.

(35) Jiang, H.; Staeglich, B.; Knoch, J.; Kumar, S.; Dilbaghi, N.; Deep, A.; Ingebrandt, S.; Pachauri, V. Programming Layer-by-Layer Liquid Phase Epitaxy in Microfluidics for Realizing Two-Dimensional Metal–Organic Framework Sensor Arrays. *Environ. Sci.: Nano* **2025**, *12* (3), 1849–1857. https://doi.org/10.1039/D4EN00764F.

(36) Fischer, J. C.; Steentjes, R.; Chen, D.-H.; Richards, B. S.; Zojer, E.; Wöll, C.; Howard, I. A. Determining Structures of Layer-by-Layer Spin-Coated Zinc Dicarboxylate-Based Metal-Organic Thin Films. *Chemistry – A European Journal* **2024**, *30* (37), e202400565. https://doi.org/10.1002/chem.202400565.

(37) Prue, A. A.; Chalmers, A. T.; Anderson, H. C.; Ralph, A. D.; Smith, S. J.; Stowers, K. J. Flexible and Rapid Synthesis of Bimetallic Metal–Organic Framework Thin Films. *Crystal Growth & Design* **2025**, *25* (9), 2849–2856. https://doi.org/10.1021/acs.cgd.4c01461.

(38) Kitagawa, S.; Kitaura, R.; Noro, S. Functional Porous Coordination Polymers. *Angewandte Chemie International Edition* **2004**, *43* (18), 2334–2375. https://doi.org/10.1002/anie.200300610.

(39) Liu, J.; Schüpbach, B.; Bashir, A.; Shekhah, O.; Nefedov, A.; Kind, M.; Terfort, A.; Wöll, C. Structural Characterization of Self-Assembled Monolayers of Pyridine-Terminated Thiolates on Gold. *Phys. Chem. Chem. Phys.* **2010**, *12* (17), 4459. https://doi.org/10.1039/b924246p.

(40) Zhuang, J.-L.; Kind, M.; Grytz, C. M.; Farr, F.; Diefenbach, M.; Tussupbayev, S.; Holthausen, M. C.; Terfort, A. Insight into the Oriented Growth of Surface-Attached Metal–Organic Frameworks: Surface Functionality, Deposition Temperature, and First Layer Order. *J. Am. Chem. Soc.* **2015**, *137* (25), 8237–8243. https://doi.org/10.1021/jacs.5b03948.

(41) Shekhah, O. Layer-by-Layer Method for the Synthesis and Growth of Surface Mounted Metal-Organic Frameworks (SURMOFs). *Materials* **2010**, *3* (2), 1302–1315. https://doi.org/10.3390/ma3021302.

(42) Liu, J.; Wöll, C. Surface-Supported Metal–Organic Framework Thin Films: Fabrication Methods, Applications, and Challenges. *Chem. Soc. Rev.* **2017**, *46* (19), 5730–5770. https://doi.org/10.1039/C7CS00315C.

(43) Srisombat, L.; Jamison, A. C.; Lee, T. R. Stability: A Key Issue for Self-Assembled Monolayers on Gold as Thin-Film Coatings and Nanoparticle Protectants. *Colloids and Surfaces A: Physicochemical and Engineering Aspects* **2011**, *390* (1–3), 1–19. https://doi.org/10.1016/j.colsurfa.2011.09.020.

(44) Willey, T. M.; Vance, A. L.; Van Buuren, T.; Bostedt, C.; Terminello, L. J.; Fadley, C. S. Rapid Degradation of Alkanethiol-Based Self-Assembled Monolayers on Gold in Ambient Laboratory Conditions. *Surface Science* **2005**, *576* (1–3), 188–196. https://doi.org/10.1016/j.susc.2004.12.022.

(45) Jones, J. A.; Qin, L. A.; Meyerson, H.; Kwon, I. K.; Matsuda, T.; Anderson, J. M. Instability of Self-assembled Monolayers as a Model Material System for Macrophage/FBGC Cellular Behavior. *J Biomedical Materials Res* **2008**, *86A* (1), 261–268. https://doi.org/10.1002/jbm.a.31660.

(46) Brewer, N. J.; Janusz, S.; Critchley, K.; Evans, S. D.; Leggett, G. J. Photooxidation of Self-Assembled Monolayers by Exposure to Light of Wavelength 254 Nm: A Static SIMS Study. *J. Phys. Chem. B* **2005**, *109* (22), 11247–11256. https://doi.org/10.1021/jp0443299.

(47) Singh, M.; Kaur, N.; Comini, E. The Role of Self-Assembled Monolayers in Electronic Devices. *J. Mater. Chem. C* **2020**, *8* (12), 3938–3955. https://doi.org/10.1039/D0TC00388C.

(48) Yang, E.; Song, X.-C.; Zhu, J.-W. 1,4-Diazoniabicyclo[2.2.2]Octane Terephthalate. *Acta Crystallogr E Struct Rep Online* **2008**, *64* (9), o1764–o1764. https://doi.org/10.1107/S1600536808025312.

(49) Gattorno, G. R.; Oskam, G. Forced Hydrolysis vs Self-Hydrolysis of Zinc Acetate in Ethanol and Iso-Butanol. *ECS Trans.* **2006**, *3* (9), 23–28. https://doi.org/10.1149/1.2357093.

(50) Kirlikovali, K. O.; Hanna, S. L.; Son, F. A.; Farha, O. K. Back to the Basics: Developing Advanced Metal–Organic Frameworks Using Fundamental Chemistry Concepts. *ACS Nanosci. Au* **2023**, *3* (1), 37–45. https://doi.org/10.1021/acsnanoscienceau.2c00046.

(51) Qi, X.; Xie, Y.; Niu, J.; Zhao, J.; Li, Y.; Fang, W.; Zhang, J. Application of Hard and Soft Acid-base Theory to Construct Heterometallic Materials with Metal-oxo Clusters. *Angew Chem Int Ed* **2025**, *64* (1), e202417548. https://doi.org/10.1002/anie.202417548.

(52) Zheng, S.; Sun, Y.; Xue, H.; Braunstein, P.; Huang, W.; Pang, H. Dual-Ligand and Hard-Soft-Acid-Base Strategies to Optimize Metal-Organic Framework Nanocrystals for Stable Electrochemical Cycling Performance. *National Science Review* **2022**, *9* (7), nwab197. https://doi.org/10.1093/nsr/nwab197.

(53) Gao, Q.; Xie, Y.-B.; Li, J.-R.; Yuan, D.-Q.; Yakovenko, A. A.; Sun, J.-H.; Zhou, H.-C. Tuning the Formations of Metal–Organic Frameworks by Modification of Ratio of Reactant, Acidity of Reaction System, and Use of a Secondary Ligand. *Crystal Growth & Design* **2012**, *12* (1), 281–288. https://doi.org/10.1021/cg201059d.

(54) Dybtsev, D. N.; Chun, H.; Kim, K. Rigid and Flexible: A Highly Porous Metal–Organic Framework with Unusual Guest-Dependent Dynamic Behavior. *Angew Chem Int Ed* **2004**, *43* (38), 5033–5036. https://doi.org/10.1002/anie.200460712.

(55) McCarthy, B. D.; Liseev, T.; Beiler, A. M.; Materna, K. L.; Ott, S. Facile Orientational Control of $M_2$ $L_2$ P SURMOFs on ⟨100⟩ Silicon Substrates and Growth Mechanism Insights for Defective MOFs. *ACS Appl. Mater. Interfaces* **2019**, *11* (41), 38294–38302. https://doi.org/10.1021/acsami.9b12407.

(56) Brandner, L. A.; Linares-Moreau, M.; Zhou, G.; Amenitsch, H.; Dal Zilio, S.; Huang, Z.; Doonan, C.; Falcaro, P. Water Sensitivity of Heteroepitaxial Cu-MOF Films: Dissolution and Re-Crystallization of 3D-Oriented MOF Superstructures. *Chem. Sci.* **2023**, *14* (43), 12056–12067. https://doi.org/10.1039/D3SC04135B.

(57) Fischer, J. C.; Li, C.; Hamer, S.; Heinke, L.; Herges, R.; Richards, B. S.; Howard, I. A. GIWAXS Characterization of Metal–Organic Framework Thin Films and Heterostructures: Quantifying Structure and Orientation. *Adv Materials Inter* **2023**, *10* (11), 2202259. https://doi.org/10.1002/admi.202202259.

(58) Oesinghaus, L.; Schlipf, J.; Giesbrecht, N.; Song, L.; Hu, Y.; Bein, T.; Docampo, P.; Müller-Buschbaum, P. Toward Tailored Film Morphologies: The Origin of Crystal Orientation in Hybrid Perovskite Thin Films. *Adv Materials Inter* **2016**, *3* (19), 1600403. https://doi.org/10.1002/admi.201600403.

# Automated Spin-Assisted Layer-by-Layer Epitaxy Produces Highly Oriented Mixed-Linker MOF Thin Films

## SUPPLEMENTARY INFORMATION

Eleonora Afanasenko,[1] Benedetta Marmiroli,[2] Behnaz Abbasgholi-NA,[3] Barbara Sartori[2], Giovanni Birarda,[4] Chiaramaria Stani,[4] Matjaž Finšgar,[5] Peter E. Hartmann,[6] Mark Bieber,[7] Emma Walitsch,[7] Rolf Breinbauer,[7] Simone Dal Zilio,[3] Sumea Klokic[2*] and Heinz Amenitsch[2*]

1 CERIC-ERIC, Trieste, Italy
2 Institute of Inorganic Chemistry, University of Technology, Graz, Austria
3 Istituto Officina dei Materiali - CNR-IOM, Trieste, Italy
4 Elettra Sincrotrone Trieste, Trieste, Italy
5 Faculty of chemistry and chemical engineering, University of Maribor, Slovenia
6 Institute of Physical and Theoretical Chemistry University of Graz, Austria
7 Institute of Organic Chemistry, University of Technology, Graz, Austria

## 1. Experiment

### 1.1. Materials and methods

Reagents and solvents are available commercially and were used as received without any further purification. The linkers benzene-1,4-dicarboxylic acid (BDC), 1,4-Diazabicyclo[2.2.2]octane (DABCO) as well as the 16-Mercaptohexadecanoic acid for the SAM were purchased from TCI Chemicals, while zinc acetate dihydrate was obtained from TCI Chemicals, absolute ethanol (EtOH, 99.8%) was bought from VWR Chemicals. Gold-deposited substrates (polycrystalline gold with thickness on Si substrate precoated with titanium, roughness < 1nm) were purchased from Georg Albert PVD-Beschichtungen. The spin-coating step was performed using a Laurell spinner (Model WS-650MZ-23NPPB), Norm-Ject 20 mL plastic syringes with a standard diameter of 19.4 mm, and programmable syringe pumps (Harvard PHD 4400). Experiments were performed at ambient conditions.

### 1.2. Synthesis of the $Zn_2BDC_2DABCO$ thin films

*Self-assembled monolayers* (SAMs) were formed on gold substrates sized 1x1cm by immersion in an ethanolic (absolute) 0.02 mM solution of 16-mercaptohexadecanoic acid (MHDA) containing 10 vol% acetic acid, or in a 0.02 mM solution of (4-(4-pyridyl)phenyl)methanethiol (PP1), for 24 hours.[1] Prior to functionalization, the wafers were sonicated in absolute ethanol for 30 minutes to ensure thorough cleaning.

The linker working solution was prepared by dissolving 0.01 mmol of benzene-1,4-dicarboxylic acid (BDC) in 100 mL of ethanol and sonicating for 2 hours to ensure complete dissolution. After confirming the proper BDC concentration *via* UV-Vis spectroscopy, 0.03 mmol of 1,4-diazabicyclo[2.2.2]octane (DABCO) was added, and the solution was sonicated for an additional hour. The working metal solution was prepared by dissolving 0.01 mmol of $Zn(CH_3CO_2)_2 \cdot 2H_2O$ in 100 mL of ethanol, followed by one hour of sonication.

## 1.3. Synthesis of the 4-(4-pyridyl)phenyl)methanethiol (PP1)

### 1) (4-Bromophenyl)methanethiol

17.0 g 4-bromobenzylbromide (68.0 mmol, 1 equiv.) were dissolved in MeOH (150 mL). Afterwards, 8.55 g potassium thioacetate (74.8 mmol, 1.1 equiv.) were added and the resulting suspension was stirred at r.t. for 30 min in an argon-flushed three-necked round-bottom flask, equipped with a Teflon-coated magnetic stirring bar. After TLC and GC-MS indicated full conversion, 9.41 g $K_2CO_3$ (68.1 mmol, 1 equiv.) were added and the suspension was stirred for another 40 min at r.t. After complete consumption of the intermediate thioacetate (reaction monitoring via TLC and GC-MS), the reaction mixture was carefully acidified (pH = 6) with 1M aq. HCl and extracted with DCM (3 x 120 mL). The combined organic phases were dried over $Na_2SO_4$, filtered and concentrated under reduced pressure. The crude product was used in the next step without further purification.

**Yield**: 13.8 g (quant.), orange oil, $C_7H_7BrS$ [203.10 g/mol].

**TLC:** $R_f$ = 0.75 (cyclohexane:EtOAc = 9:1; UV).

**GC-MS**: (EI, 70 eV; MT_50_S): $t_R$ = 5.22 min; *m/z* (%): 204 (26), 169 (100), 121 (6), 90 (30), 63 (12).

**$^1$H NMR** (300 MHz, $CDCl_3$): δ 7.44 (d, *J* = 8.3 Hz, 2H), 7.20 (d, *J* = 8.3 Hz, 2H), 3.69 (d, *J* = 7.6 Hz, 2H), 1.76 (t, *J* = 7.6 Hz, 1H; -S**H**) ppm.

**$^{13}$C NMR** (76 MHz, $CDCl_3$): δ 140.2, 131.9, 129.9, 121.0, 28.5 ppm.

### 2) ((4-Bromobenzyl)thio)triisopropylsilane

An argon-flushed Schlenk flask, equipped with a Teflon-coated magnetic stirring bar, was charged with 11.6 g (4-bromophenyl)methanethiol (57.1 mmol, 1 equiv.) and dissolved in abs. THF (110 mL). The reaction mixture was cooled to 0 °C and 2.51 g NaH (62.8 mmol, 1.1 equiv.) were added in portions over 15 min. The Schlenk flask was equipped with a bubbler and the reaction mixture was stirred until the evolution of gas had ceased (50 min). Afterwards, 13 mL TIPSCl (62.8 mmol, 1.1 equiv.) were added dropwise over 15 min and the reaction mixture was stirred for 70 min at r.t. (reaction monitoring via GC-MS). Subsequently, the reaction was quenched by the addition of $H_2O$ (110 mL). The phases were separated and the aqueous phase was back-extracted with EtOAc (3 x 110 mL). The combined organic phases were dried over $Na_2SO_4$, filtered and concentrated under reduced pressure. The crude product was first purified via vacuum distillation (10.0 x 2.0 com Vigreux column, 0.041 mbar) and then via flash column chromatography (1 L $SiO_2$, 26.0 x 8.0 cm, cyclohexane to cyclohexane:EtOAc = 10:1 (v/v)).

**Yield**: 13.5 g (66%), colourless liquid, $C_{16}H_{27}BrSSi$ [359.44 g/mol].

**TLC:** $R_f$ = 0.39 (cyclohexane; UV).

**b.p:** 119-120 °C (0.041 mbar).

**GC-MS**: (EI, 70 eV; MT_50_S): $t_R$ = 7.69 min; *m/z* (%): 360 (1), 315 (50), 273 (7), 169 (100), 90 (24).

**$^1$H NMR** (300 MHz, $CDCl_3$): δ 7.42 (d, *J* = 8.3 Hz, 2H), 7.22 (d, *J* = 8.3 Hz, 2H), 3.69 (s, 2H), 1.30 (hept, *J* = 7.1 Hz, 3H), 1.13 (d, *J* = 7.1 Hz, 18H) ppm.

**$^{13}$C NMR** (76 MHz, $CDCl_3$): δ 140.0, 131.7, 130.3, 120.7, 29.7, 18.71, 12.9 ppm.

### 3) 4-(4-(((Triisopropylsilyl)thio)methyl)phenyl)pyridine

An argon-flushed Schlenk flask, equipped with a Teflon-coated magnetic stirring bar, was charged with 1.67 g 4-pyridylboronic acid (90% purity, 12.2 mmol, 1.2 equiv.), 6.67 g $CsCO_3$ (20.5 mmol, 2 equiv.), 374 mg Pd(dppf)$Cl_2$ (0.512 mmol, 0.05 equiv.), 370 µL $H_2O$ (20.5 mmol, 2 equiv.) and 3.66 g ((4-bromobenzyl)thio)triisopropylsilane (10.2 mmol, 1 equiv.) and dissolved in degassed 1,4-dioxane (30 mL). Then, the reaction mixture was stirred at 90 °C for 20 h. After complete consumption of the starting material (reaction monitoring via GC-MS), the reaction mixture was cooled to r.t.. The mixture was taken up in $H_2O$ (30 mL) and extracted with EtOAc (4 x 30 mL). The combined organic phases were dried over $Na_2SO_4$, filtered and concentrated under reduced pressure. The crude product was purified via flash column chromatography (500 mL $SiO_2$, 20.0 x 6.0 cm, cyclohexane:EtOAc = 4:1 (v/v) to cyclohexane:EtOAc = 1:1 (v/v)).

**Yield**: 2.99 g (83%), beige, waxy solid, $C_{21}H_{31}NSSi$ [357.63 g/mol].

**TLC:** $R_f$ = 0.45 (cyclohexane:EtOAc = 1:1; UV, $KMnO_4$).

**GC-MS**: (EI, 70 eV; MT_50_S): $t_R$ = 9.80 min; *m/z* (%): 357 (10), 314 (49), 168 (100), 139 (3), 115 (5).

**$^1$H NMR** (300 MHz, $CD_2Cl_2$): δ 8.68 – 8.57 (m, 2H), 7.62 (d, *J* = 8.1 Hz, 2H), 7.55 – 7.43 (m, 4H), 3.81 (s, 2H), 1.40 – 1.25 (m, 3H), 1.16 (d, *J* = 7.2 Hz, 18H) ppm.

**$^{13}$C NMR** (76 MHz, $CD_2Cl_2$): δ 150.8, 148.3, 142.8, 137.2, 129.8, 127.6, 121.9, 30.3, 18.9, 13.3 ppm.

### 4) (4-(Pyridin-4-yl)phenyl)methanethiol

Two batches have been prepared consecutively as described in the following procedure:

In an argon-flushed 500 mL two-necked round-bottom flask, equipped with a Teflon-coated magnetic stirring bar, a mixture of MeOH (150 mL) and 6M aq. HCl (30 mL) were degassed by passing a stream of nitrogen through the solution for 1 h. Afterwards, 2.60 g 4-(4-(((triisopropylsilyl)thio)methyl)phenyl)pyridine (7.27 mmol, 1 equiv.) were added and the mixture was heated under reflux for 20 h. The reaction mixture was allowed to cool to r.t. and diluted with a mixture of MeOH and $H_2O$ (150 mL; 1:1 (v/v)). The mixture was washed with *n*-pentane (1 x 50 mL, 2 x 100 mL) to remove apolar impurities before the MeOH was removed under reduced pressure. The resulting aqueous solution was again degassed by sonication and passing a stream of nitrogen through the solution for 15 min. Afterwards, the pH of the solution was adjusted to pH ~6 by the addition of solid trisodium citrate dihydrate. The slurry was immediately extracted with DCM (3 x 100 mL). The combined organic phases were dried over $Na_2SO_4$, filtered and concentrated under reduced pressure. The crude product was dissolved in boiling, degassed methyl cyclohexane (250 mL), filtered hot through a Schlenk frit and left for crystallization. The crystals were collected by filtration, washed with *n*-pentane and dried *in vacuo*. Further purification was performed by evaporation and resublimation on a cooling finger (105 °C, 0.15 mbar).

**Yield**: 2.09 g (53% (2 batches)), colourless solid, $C_{12}H_{11}NS$ [201.29 g/mol].

**GC-MS**: (EI, 70 eV; MT_50_S): $t_R$ = 6.97 min; *m/z* (%): 201 (18), 168 (100), 139 (8), 63 (5).

**m.p.** = 62-64 °C.

**$^1$H NMR** (400 MHz, $CDCl_3$): δ 8.67 – 8.62 (m, 2H), 7.62 – 7.57 (m, 2H), 7.52 – 7.47 (m, 2H), 7.46 – 7.41 (m, 2H), 3.79 (d, *J* = 7.6 Hz, 2H), 1.81 (t, *J* = 7.6 Hz, 1H; -S**H**) ppm.

**$^{13}$C NMR** (101 MHz, $CDCl_3$): δ 150.3, 148.0, 142.4, 136.9, 128.9, 127.4, 121.6, 28.7 ppm.

## 1.2. Characterization techniques

### 1.2.1. Grazing Incidence Small Angle X-Ray Scattering (GIWAXS) Experiments

GIWAXS measurements of the synthesized samples prepared under different stoichiometric conditions were performed at the Austrian SAXS beamline at ELETTRA, Trieste. An X-ray beam with a wavelength of 1.54 Å (beam energy 8 keV) was used with a sample-to-detector distance of 260 mm, providing a $q$-range of $0.01 < q < 29.21$ $nm^{-1}$, where $q$ denotes the magnitude of the scattering vector. The beam size was 0.1 × 1 mm (v × h). The angular scale of the detector was calibrated using silver behenate. Samples were mounted on a motorized stage with a resolution of 0.001°, and the incident grazing angle was set to 1.6°.

The 2D GIWAXS pattern were acquired in triplicates with an exposure time of 30 s each. A Pilatus3 1M detector (Dectris Ltd., Baden, Switzerland; active area 169 × 179 $mm^2$, pixel size 172 µm) was used, and the detector images were processed using SAXSDOG, a software developed at the Austrian SAXS beamline for automatic data reduction.[2] For data analysis, radially integrated diffraction patterns, out-of-plane and in-plane cuts were considered. The cut margin for the out-of-plane and in-plane integration was set to 10 pixels. The integrated data were further processed using IGOR Pro (Wavemetrics, Inc., Lake Oswego, OR). Crystalline phases were identified by comparing the Bragg peaks positions of the respective samples with bulk diffraction patterns of the guest-free single-crystal DMOF structure (see Figure S6).[3]

### 1.2.2. IR spectromicroscopy

Infrared measurements were performed at the SISSI-Bio offline end station of the SISSI beamline at Elettra Sincrotrone Trieste. The infrared spectromicroscopy setup is equipped with a grazing-angle objective (GI-IR), which is essential for the analysis of thin films supported on metallic or highly reflective substrates. Spectra were acquired in slow acquisition mode by averaging 128 scans at a spectral resolution of 2 $cm^{-1}$ and a scanner speed of 60 kHz. The aperture were set to 75 × 75 $\mu m^2$ to achieve an optimal signal-to-noise ratio and enable detection of subtle spectral variations. A broadband mercury cadmium telluride detector (MCT-B), operating in the 3800–400 $cm^{-1}$ spectral range, was used for signal detection. All the collected spectra were pre-processed with Bruker software OPUS 9.7. The corresponding background reference spectrum was manually subtracted for minimizing the contribution of water and $CO_2$ to the final spectra.

### 1.2.3. SEM and EDS measurements

Morphologies of samples were investigated by a scanning electron microscope (FIB–SEM; Aquilos2, Thermo Fisher Scientific, The Netherlands) and cross-sectional analyses were carried out using a focused ion beam in combination with scanning electron microscopy (FIB–SEM; Helios G5 UC, Thermo Fisher Scientific, The Netherlands) equipped with a multi gas injection system (GIS) and an Ultim Max 65 energy-dispersive X-ray spectroscopy (EDS) detector (Oxford Instruments, UK). Prior to analysis, the samples were sputter-coated with a 6 nm platinum (Pt) layer using a PECS 682 system (Gatan, USA) to minimize surface charging.

### 1.2.4. ToF-SIMS measurements

Time-of-flight secondary ion mass spectrometry (ToF-SIMS) measurements were performed using a M6 instrument (IONTOF GmbH, Münster, Germany) equipped with a bismuth liquid metal ion gun (LMIG) for analysis and an argon gas cluster ion beam (GCIB) for sputtering. Measurements were conducted in negative ion mode. For spectral acquisition and imaging, a pulsed $Bi_3^+$ primary ion beam was employed as the analysis source. Depth profiling experiments were carried out using an $Ar_{2000}^+$ GCIB operated at an acceleration energy of 2.5 keV and a sputter current of 1 nA. The sputter beam was rastered over an area of 500 µm by 500 µm, while secondary ions were collected only from the central analysis region over an area of 300 µm by 300 µm (128 by 128 pixels) to avoid contributions from crater edges. Analysis and sputtering were alternated in a cyclic manner to obtain depth-resolved chemical information. High lateral resolution two-dimensional chemical imaging was performed using the delayed extraction analyzer mode in combination with the fast-imaging LMIG mode. Imaging was conducted over a field of view of 52 µm by 52 µm (512 by 512 pixels). Charge compensation was applied during all measurements using a low-energy electron flood gun (20 eV). Mass calibration was performed using the signals for $C_3^-$ at mass-to-charge (*m*/*z*) 36.00, $C_4^-$ at *m*/*z* 48.00, and $C_5^-$ at *m*/*z* 60.00. All data acquisition, visualization, and post-processing were carried out using SurfaceLab 7.6 software (IONTOF GmbH).

Two-dimensional ToF-SIMS images were acquired using the delayed extraction mode combined with fast imaging LMIG mode. This approach provides high mass resolution (~10,000) while maintaining lateral resolution of ~100 nm. The coverage of $ZnOH^-$ was calculated as the fraction of pixels with intensities above a threshold value of 1 to the total number of pixels.

#### 1.2.5. 3D Profilometry

The surface topography and the analysis of the sputter crater depths formed during GCIB sputtering in the ToF-SIMS measurements were evaluated using a DektakXT stylus profilometer controlled with Visio64 software program (Bruker, Karlsruhe, Germany). The resulting 3D surface profiles were subsequently processed and evaluated with MountainsMap® Imaging Topography software (Digital Surf, Besançon, France). During post-processing, levelling was applied to correct for sample tilt and align the measured surface to a common reference plane, while form removal was used to subtract the large-scale surface shape or curvature, enabling clearer evaluation of the crater geometry and local topographic features. Three-dimensional surface topography maps were acquired over areas of 3 by 3 $mm^2$ and 1.5 by 2.5 $mm^2$.

## 2. LPE Spin-Coating Workflow

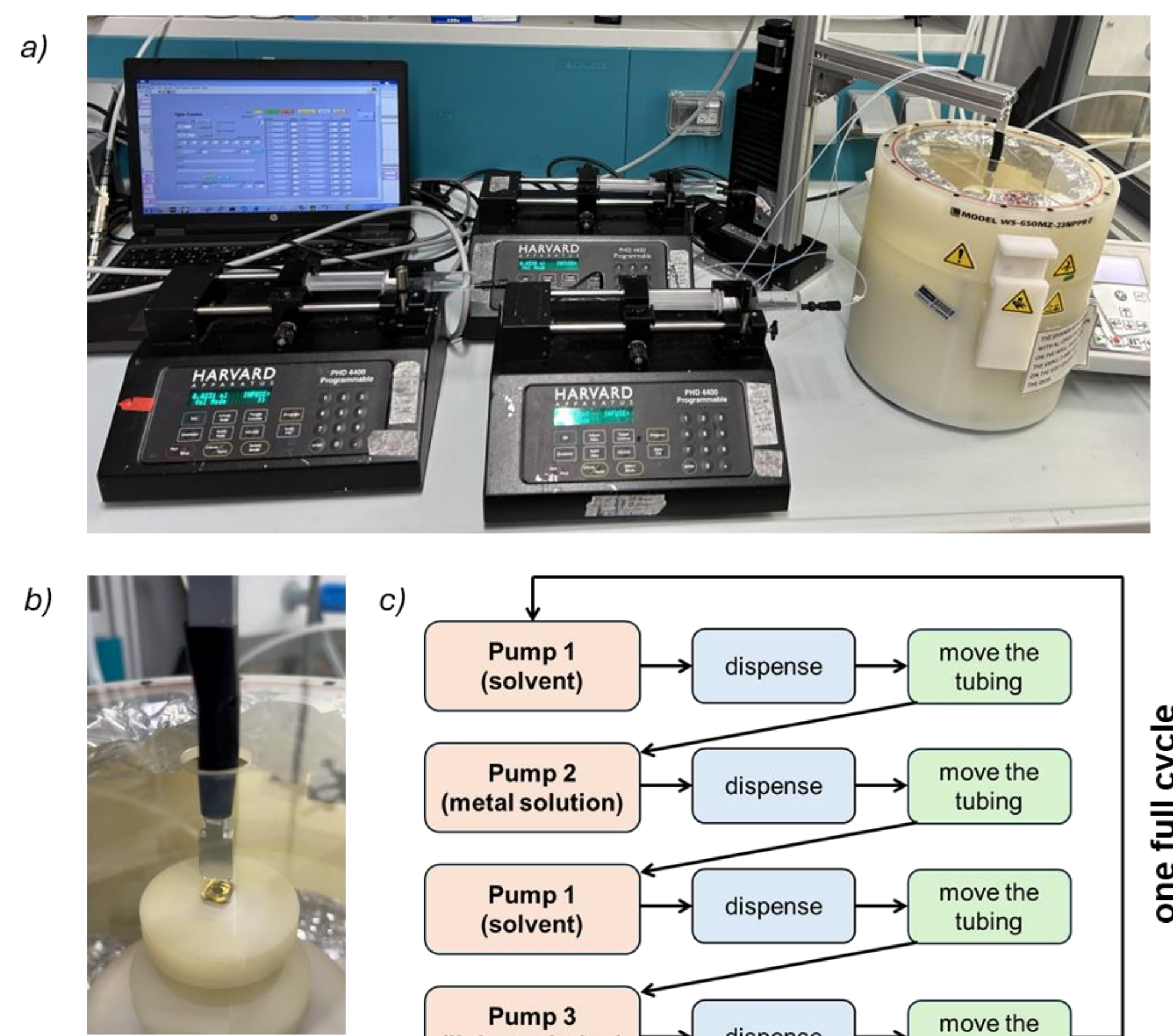

**Figure S1. Process overview of the automatic deposition step.** (a) Photograph of the spin-assisted LPE-LbL setup; (b) The tube holder moves according to the programmed sequence to precisely dispense the solution at the center of the spinned wafer. (c) Outline of the programmed sequence of steps representing one deposition cycle.

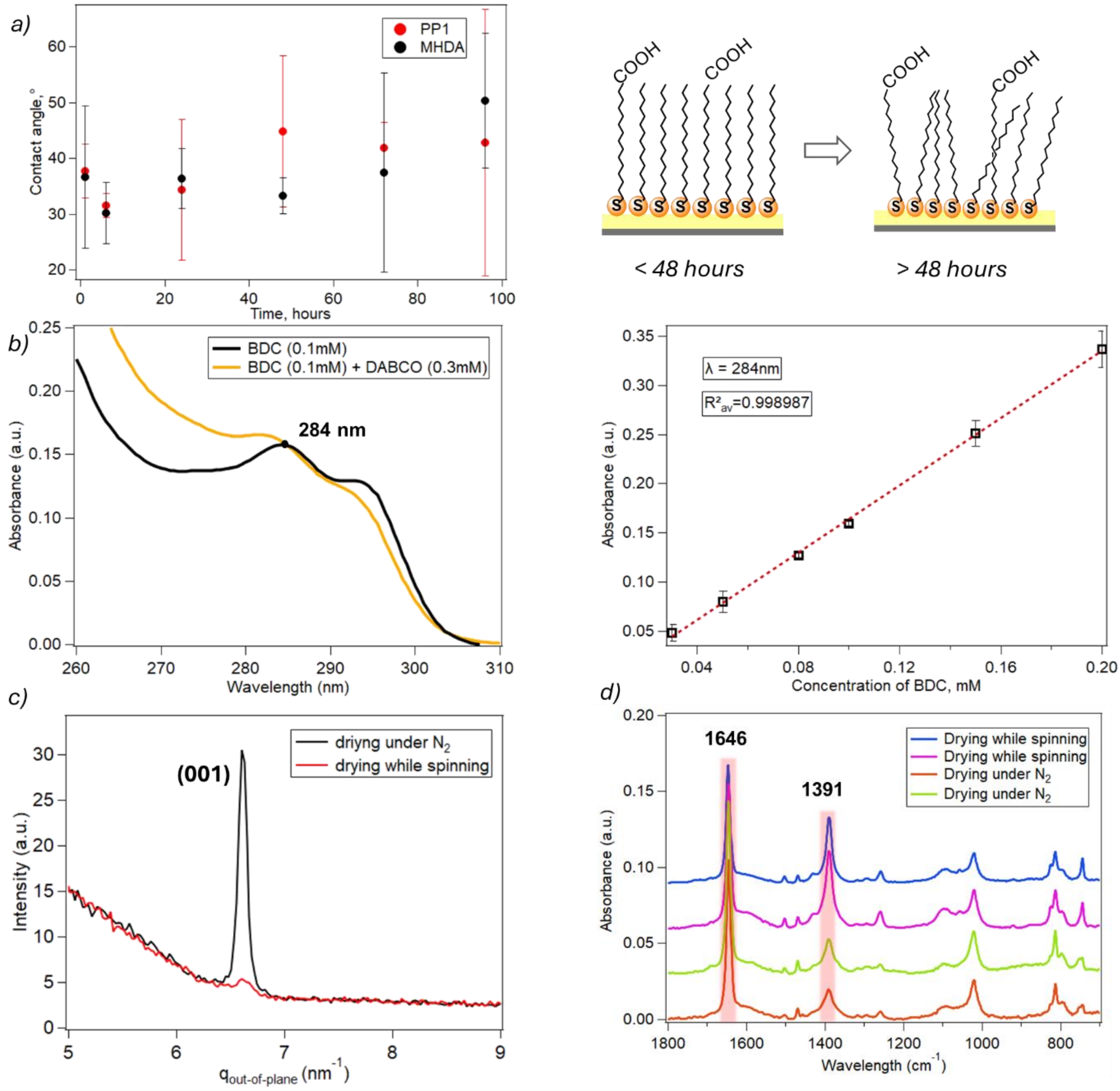

**Figure S2**. **Quality control during the fabrication process.** (a) Contact angle measurements show a progressive loss of order in the SAM layer within time. Each data point represents the average of three independent samples, and the error bands indicate the presence of structural variability even under optimal functionalization conditions.[4] However, functionalized substrates used within the first 48 h enable reproducible growth of highly oriented, homogeneous MOF films. (b) UV-Vis absorption spectra of a 0.1 mM BDC solution and the working solution after addition of DABCO at a 1:3 ratio (left). The band shift at 284 nm limits accurate concentration determination in the mixed-ligand system; therefore, only BDC was quantified using its calibration curve at this wavelength (right). (c) GIWAXS, out-of-plane cuts, illustrating the loss of crystallinity when the sample is allowed to dry during spinning as indicated by the weakening of the (001) reflection. (d) The GI-IR spectra contain characteristic

asymmetric (1646 $cm^{-1}$) and symmetric (1394 $cm^{-1}$) stretching vibrations of coordinated carboxylic groups indicating MOF formation and confirm that the chemical composition of the samples remains unchanged under different drying conditions.

## 3. Assessment of film uniformity and fabrication reproducibility

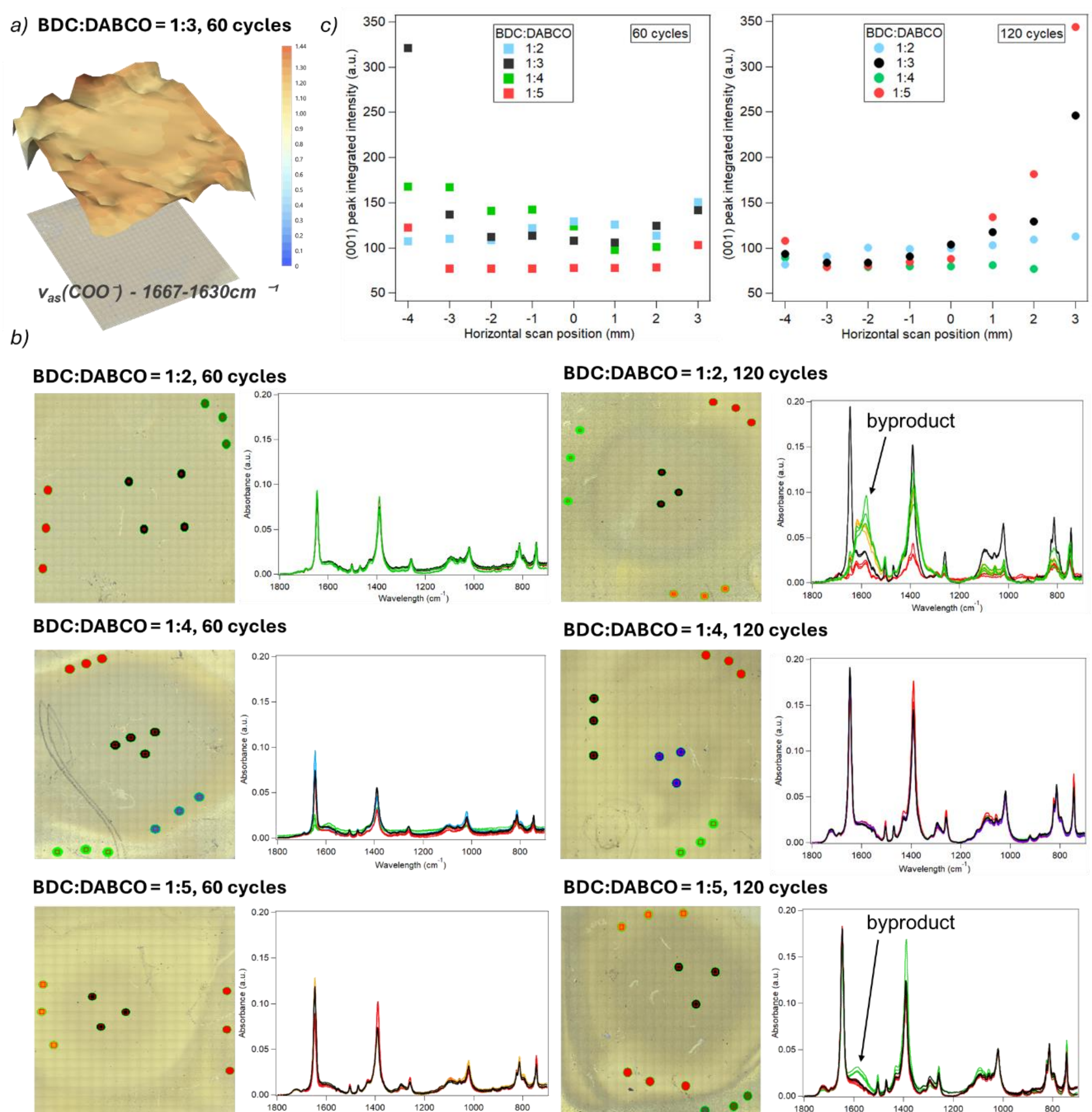

**Figure S3. IR and GIWAXS uniformity analysis.** (a) 3D projection of the intensity of the asymmetric stretching vibrations of the carboxylate group in the range of 1667-1630cm$^{-1}$ over the full sample area equal to 1 cm x 1 cm (BDC:DABCO= 1:3) with 60 cycles, indicating formation of MOF. (b) IR-data for the samples with BDC:DABCO=1:2, 1:4, 1:5 molar ratios with 60/120. To probe the homogeneity, each wafer was first imaged and subsequently probed at 3-4 different locations (center and edges) to compare spectral features across distinct surface regions. Colour of the spectra in the central column corresponds to the colour of the point in the image. Samples prepared

with 60 deposition cycles and varying reagent molar ratios exhibit high uniformity across the entire sample, including also the edges. In contrast, increasing the number of cycles to 120 results in a slight loss of homogeneity and the appearance of by-products at the wafer edges. (c) GIWAXS measurements across the sample, where 0 mm is the position at the centre of the sample. Integrated out-of-plane intensity versus horizontal positions scanning the entire sample confirm higher uniformity of the 60 cycles samples.

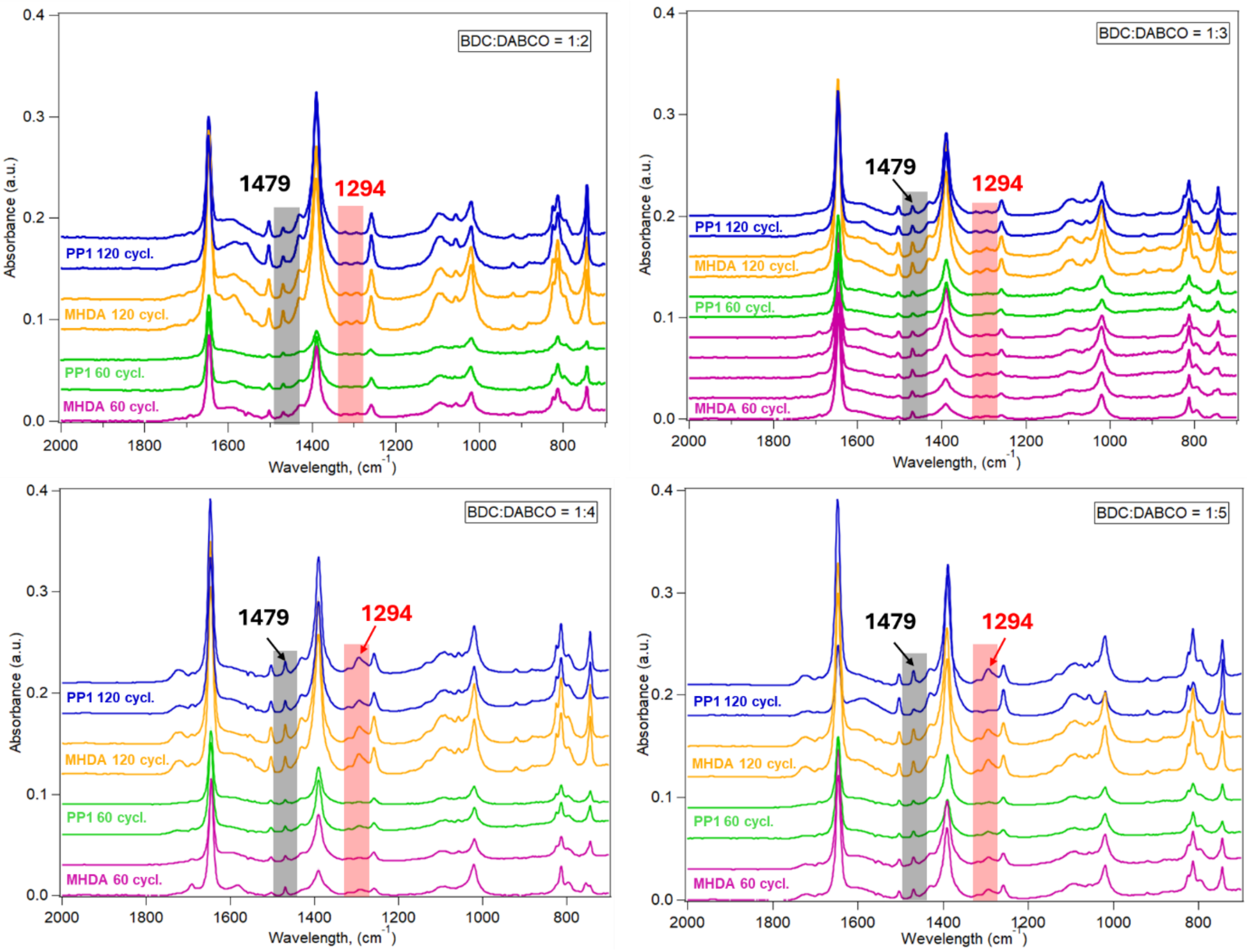

**Figure S4.** IR spectra of samples prepared with varying BDC:DABCO molar ratios and deposition cycles confirm the reproducibility of the LbL-LPE spin-coating method. The band at 1294 cm$^{-1}$, assigned to unreacted DABCO, appears only in the 1:5 samples and in the 1:4 samples after 120 cycles due to increased material deposition. In contrast, the band at 1479 cm$^{-1}$ corresponds to coordinated DABCO and is weak in samples synthesized with a 1:2 ligand ratio, 60 cycles.

## 4. DFT calculations

To reproduce the experimental spectrum as accurately as possible and enable reliable assignment of the bands to their vibrational modes, DFT calculations were performed.

### 4.1. Computational Details

The initial model structures of $Zn_2BDC_2DABCO$ were generated based on the computational structures of Griffith et al.[5] using VESTA and the GUI of the atomic simulation environment (ASE) to yield a $\sqrt{2} \times \sqrt{2} \times 1$ supercell.[6]

DFT calculations were performed with FHI-aims[7–12] (version 210716_2) employing light settings[7] and the PBE functional[13] with the many-body dispersion correction[14,15] (PBE+MBD/light). Relativistic effects were treated by the scalar atomic ZORA approximation as implemented in FHI-aims. The k-grids for both the periodic DFT calculations and the MBD energies and forces were set to always satisfy $nq > 30$ Å, with $q$ being the cell length in the respective direction and $n$ being the number of k-points in that direction. Convergence criteria were set to $10^{-6}$ eV for the energy, $10^{-4}$ eV/Å for the gradient, $10^{-3}$ for the sum of eigenvalues and $10^{-5}$ for the density.

Optimizations with fixed cell parameters were performed with the FHI-AIMS implementation of the BFGS optimizer with an energy tolerance of 0.001 eV and enforcing a maximum force of 0.005 eV/Å (Figure SI1 to SI3).
Full cell optimizations were performed by employing the BFGS implementation and ExpCellFilter class of ASE,[7] enforcing a maximum force of 0.005 eV/Å.

Vibrational spectra were calculated by the finite displacement approach as implemented in Phonopy.[16,17] The respective calculations were performed on the same $\sqrt{2} \times \sqrt{2} \times 1$ supercells with finite displacements of 0.005 Å. The optimized geometries were checked to not exhibit any imaginary modes at the $\Gamma$-point. Born effective charges were calculated *via* a finite differences approach using the BEC.py script of FHI-aims (displacements were chosen to be 0 and 0.005 Å and the polarization k-grid along the reciprocal lattice vectors was set to $k_x$ = 8 x 2 x 2, $k_y$ = 2 x 8 x 2 and $k_z$ = 2 x 2 x 8). The dielectric tensors were calculated with Density-functional perturbation theory (DFPT), retaining all other relevant settings. Infrared spectra at the individual volumes were obtained *via* the Phonopy-Spectroscopy code.[18,19]

Animation files for the Phonon modes were generated with Phonopy and visualized with Jmol.[16,17,20] Similarly, the geometries depicted below were rendered in Jmol.

## 4.2. Detailed Information on the Computations

Firstly, full cell optimizations of $Zn_2BDC_2DABCO$ were performed for five different model structures (**I** to **V**). The resulting geometries (Figures S5 to S7) exhibit only subtle structural differences, especially regarding DABCO orientation and the arrangement of the coordination sphere metal nodes (the resulting cell parameters are summarized in Table S2). The obtained volumes are in good agreement with the experimental X-ray structure (with deviations around 0.6%, see Table S1) and the electronic energies of the structures differ only marginally and are within the error of the method (the maximum absolute deviation is $2.28\times10^{-3}$ eV or 0.22 kJ/mol). Gratifyingly, the obtained IR-spectra display almost no dependence on the structural model used, as they are visually virtually indistinguishable (Figures S10).

Secondly, an alternative way to obtain calculated IR spectra that accurately reflect the experimental data and, thus, allow reliable assignment of the individual bands, is to fix the cell parameters to their experimentally determined values. Hence, in addition to the above outlined full cell optimization approach we also opted to pursue this strategy to assess the robustness of our calculated IR spectra and their dependence on the chosen computational model. Towards this, all model structures **I** to **V** were also optimized with the cell parameters fixed to the experimentally determined values (termed **I-exp.** to **V-exp.**).[5] Of the five structures, only **I-exp.** and **V-exp.** did not exhibit an imaginary frequency at the $\Gamma$-point. Their geometries are displayed and compared to their respective parent structures **I** and **V** in Figures S8 and S9. Importantly, the IR-spectra obtained for **I-exp.** and **V-exp.** only slightly differ from the ones obtained from the fully optimized model structures **I** to **V** (Figures S5- S7), rendering the qualitative conclusions obtained independent of the chosen computational approach, *i.e.*, full cell optimization versus internal cell optimization with cell parameters fixed to their experimental values. This also applies to the peak assignment, see section 5, Table S3. For the main text of the publication, the IR-spectrum obtained from model structure **V-exp.** was chosen.

**Table 1**. Cell volumes and electronic energies of the calculated model structures and comparison to the experiment (entry “Exp.”).

| structure | V [Å$^3$] | E [eV] | ΔE [eV] | ΔE vs. **I** [kJ/mol] | ΔV vs exp. [Å$^3$] | ΔV vs exp. [%] |
|---|---|---|---|---|---|---|
| **I** | 2308.236 | -281489.208635371 | 0.00 | 0.00 | 13.438 | 0.59 |
| **II** | 2309.324 | -281489.208479918 | $1.04\times10^{-4}$ | 0.01 | 14.526 | 0.63 |
| **III** | 2310.473 | -281489.207031581 | $1.55\times10^{-3}$ | 0.15 | 15.676 | 0.68 |
| **IV** | 2310.421 | -281489.206387599 | $2.28\times10^{-3}$ | 0.22 | 15.624 | 0.68 |
| **V** | 2310.590 | -281489.207266224 | $1.35\times10^{-3}$ | 0.13 | 15.792 | 0.69 |
| **I-exp.** | 2294.797 | -281489.149519445 | $5.91\times10^{-2}$ | 5.70 | 0.000 | 0.00 |
| **V-exp.** | 2294.797 | -281489.150958683 | $5.77\times10^{-2}$ | 5.57 | 0.000 | 0.00 |
| Exp. | 2294.797 | n.a. | n.a. | n.a. | 0.000 | 0.00 |

**Table 2.** Cell parameters of the calculated model structures and comparison to the experiment (entry “Exp.”).

| structure | a [Å] | b [Å] | c [Å] | α [°] | β [°] | γ [°] |
|---|---|---|---|---|---|---|
| **I** | 15.724 | 15.419 | 9.523 | 91.057 | 89.123 | 90.010 |
| **II** | 15.610 | 15.534 | 9.524 | 90.490 | 89.358 | 89.999 |
| **III** | 15.581 | 15.565 | 9.527 | 89.714 | 89.896 | 89.997 |
| **IV** | 15.634 | 15.514 | 9.526 | 90.203 | 89.781 | 90.006 |
| **V** | 15.595 | 15.551 | 9.527 | 89.761 | 89.853 | 89.998 |
| **I-exp.** | 15.455 | 15.455 | 9.608 | 90.000 | 90.000 | 90.000 |
| **V-exp.** | 15.455 | 15.455 | 9.608 | 90.000 | 90.000 | 90.000 |
| Exp. | 15.455 | 15.455 | 9.608 | 90.000 | 90.000 | 90.000 |

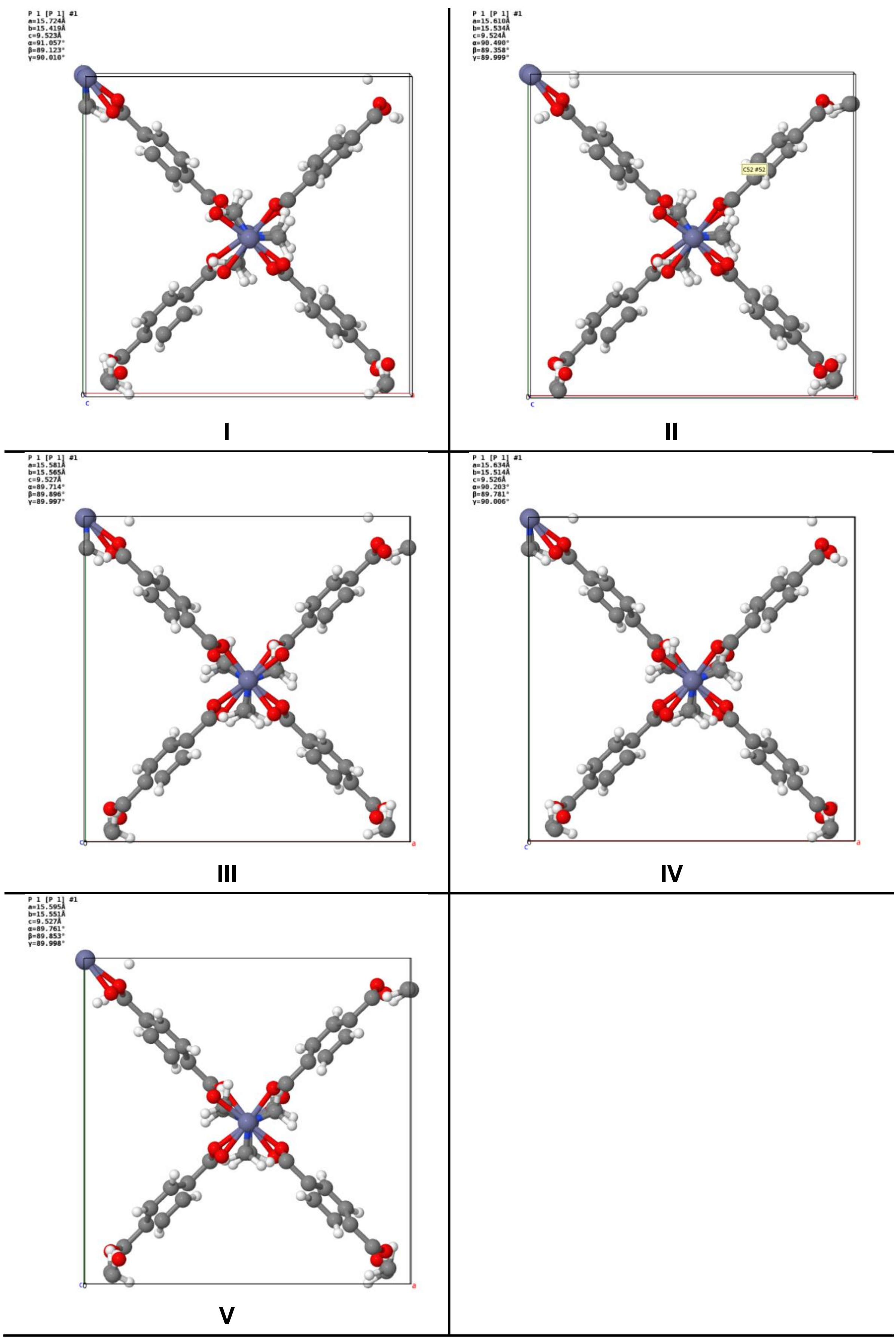

**Figure S5.** Rendered images of the 3D structures of models **I** to **V** (view along the c-axis).

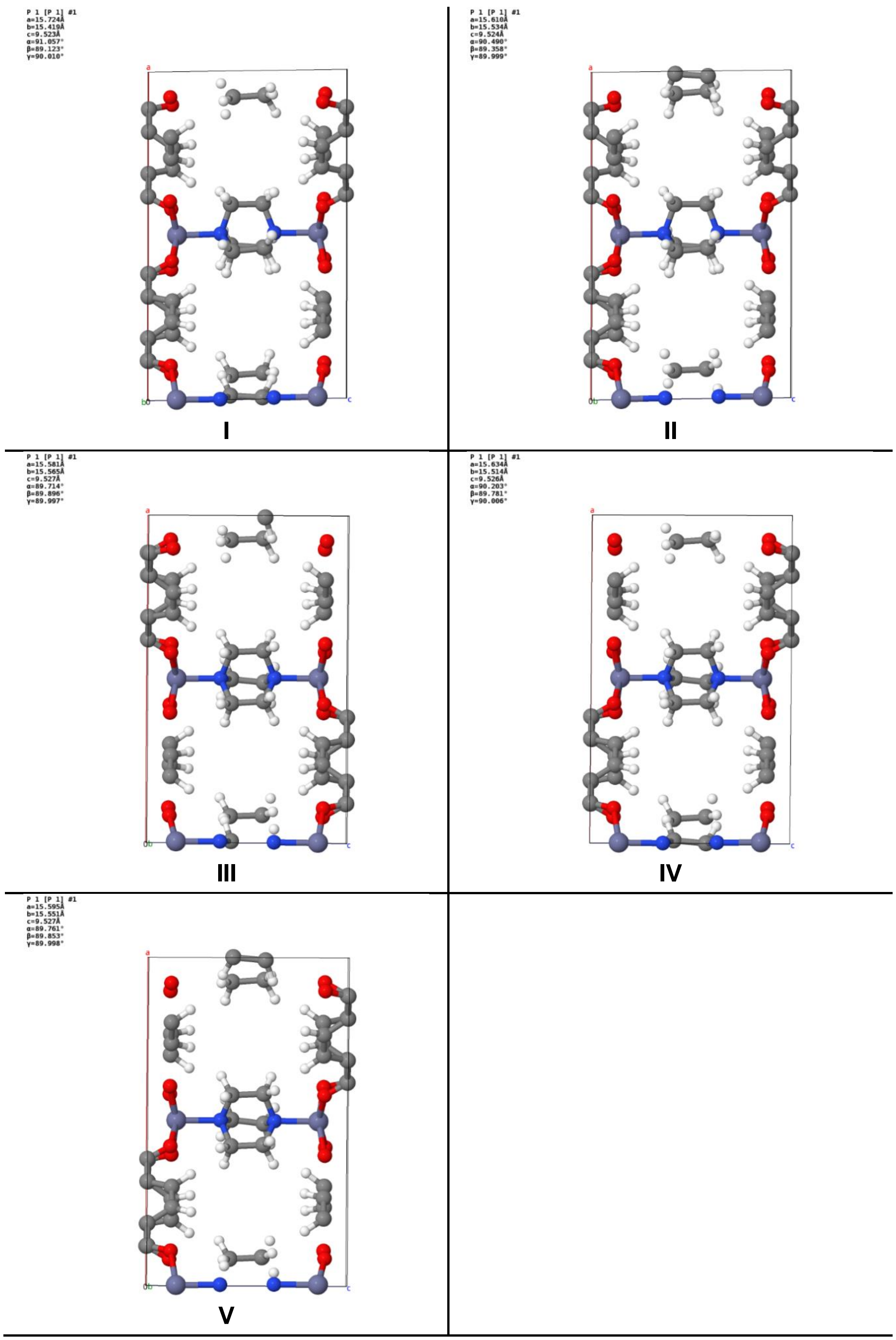

**Figure S6.** Rendered images of the 3D structures of models **I** to **V** (view along the b-axis).

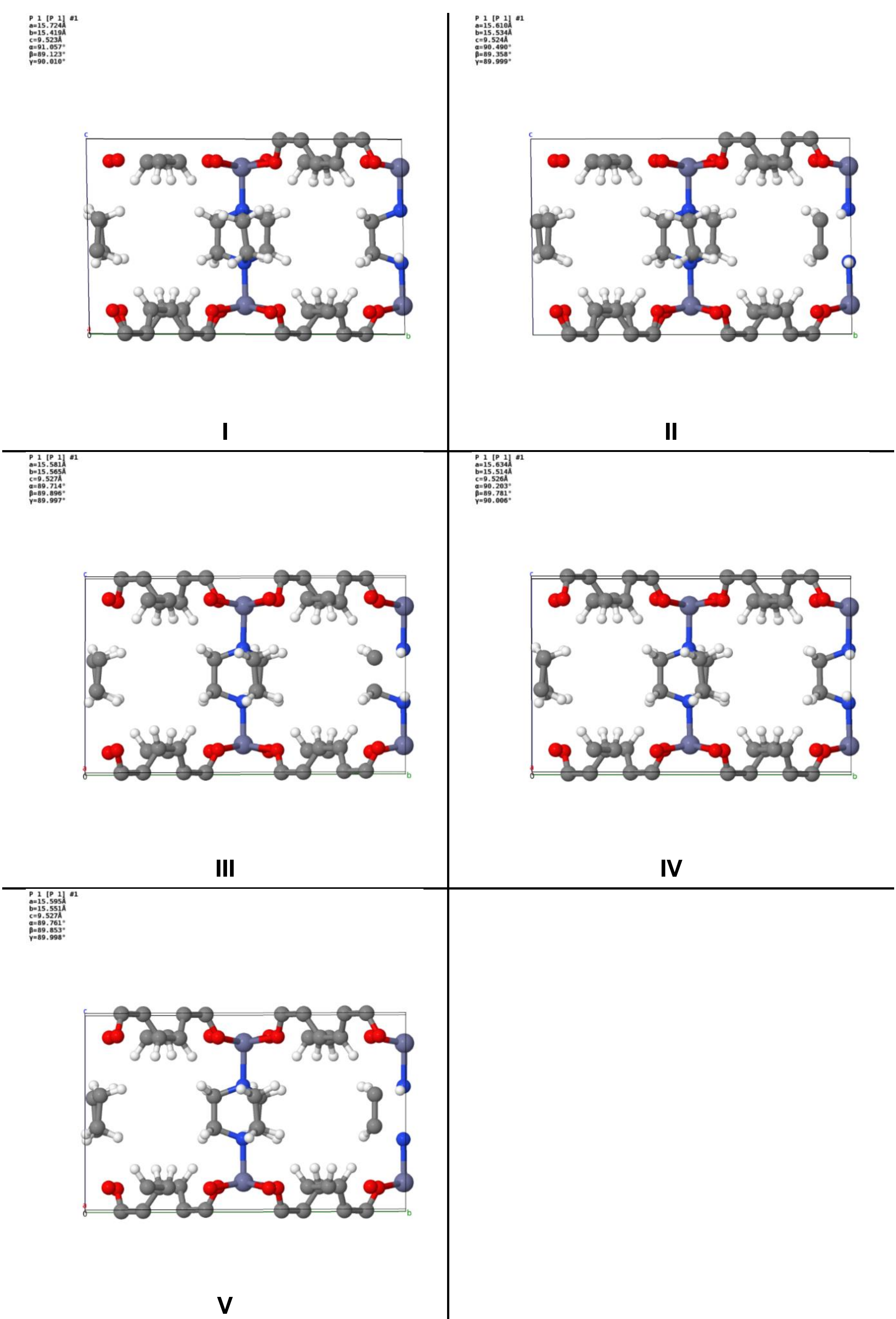

**Figure S7.** Rendered images of the 3D structures of models **I** to **V** (view along the a-axis).

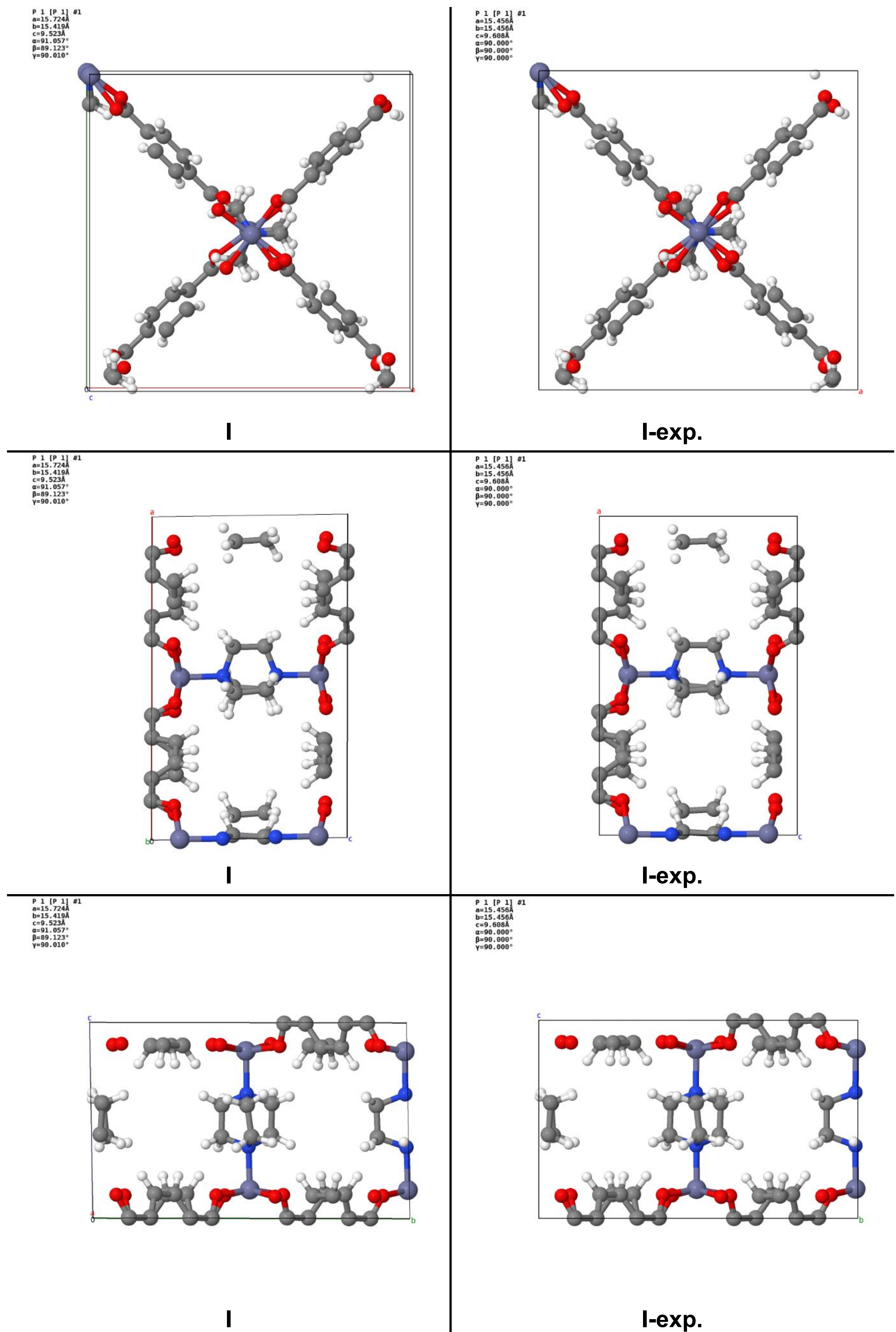

**Figure S8.** Comparison of the 3D structures of **I** (left) and **I-exp.** (right).

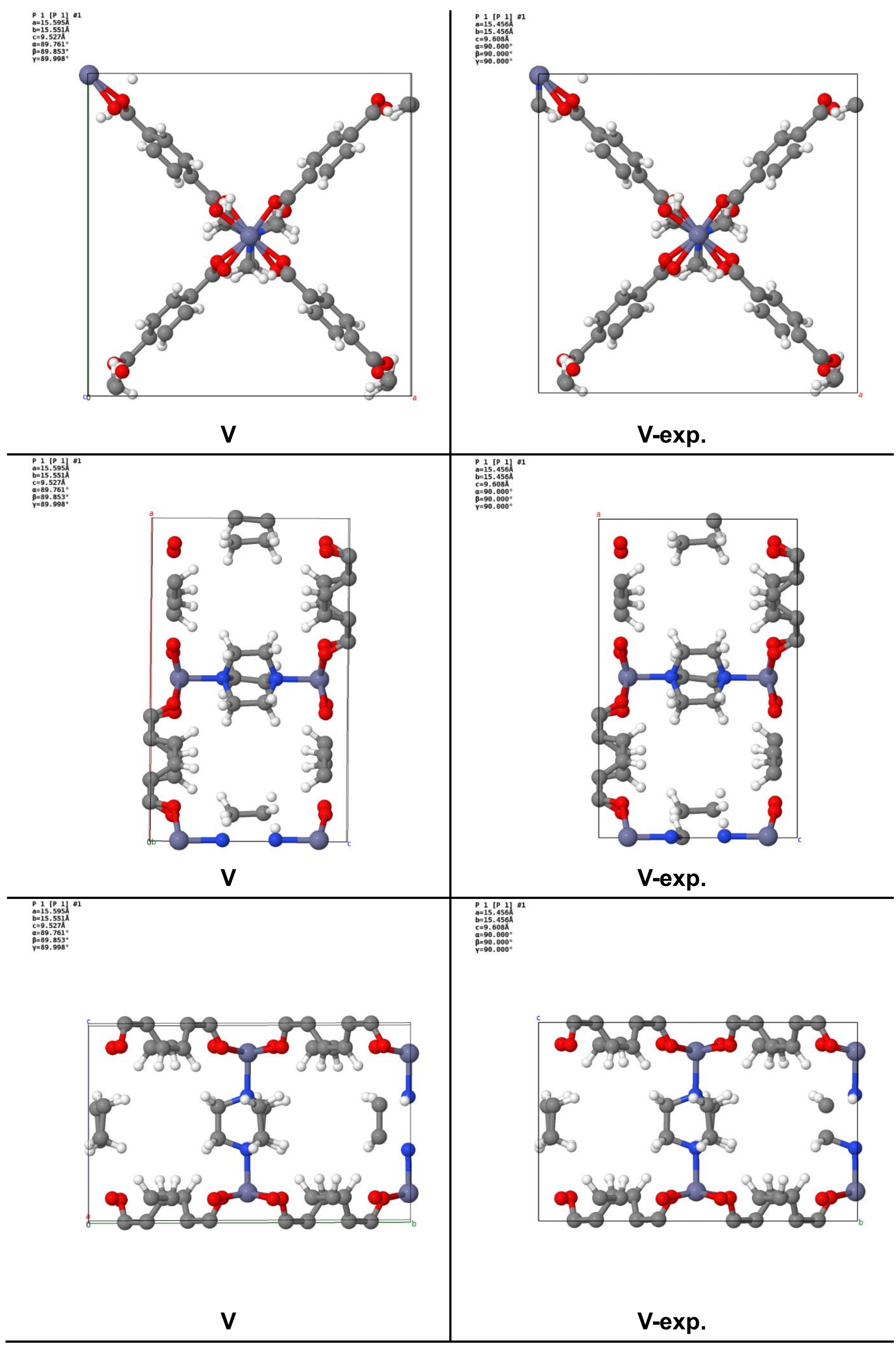

**Figure S9.** Comparison of the 3D structures of **V** (left) and **V-exp.** (right).

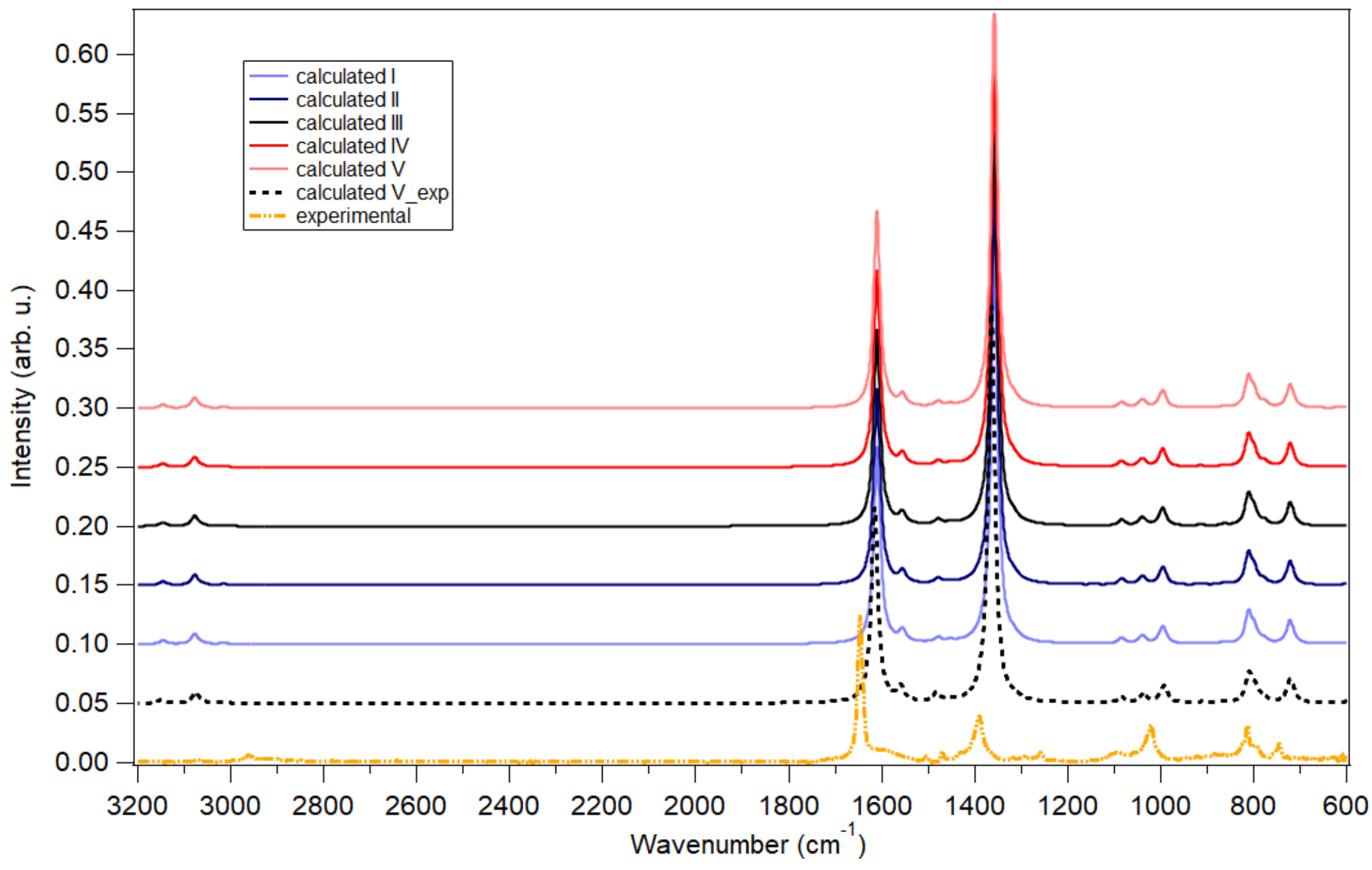

**Figure S10.** Overlay of the calculated IR-spectra for the investigated model structure.

## 5. Calculated vs. experimental IR spectra and vibrational peaks assignment

To compare the two computational model approaches and obtain a conclusion that is qualitatively independent of the chosen model, the assignment of the peaks was done on structures **IV** and **V-exp.,** representative of the two above outlined computational model approaches, i.e., full cell optimization versus internal cell optimization with cell parameters fixed to their experimental values. The computational spectra of **IV** and **V-exp.** both well reflect the experimental spectrum (Figure S11). The individual bands in the experimental spectrum were assigned to the respective type of mode by selecting the most intense individual vibrations making up the corresponding bands in the theoretical spectra of **IV** and **V-exp.**, subsequent visual inspection in the animation file and, where possible, compared to literature.[21–23] Gratifyingly, the assignment of the peaks does not differ between **IV** and **V-exp.**, rendering the obtained picture independent of the computational model approach.

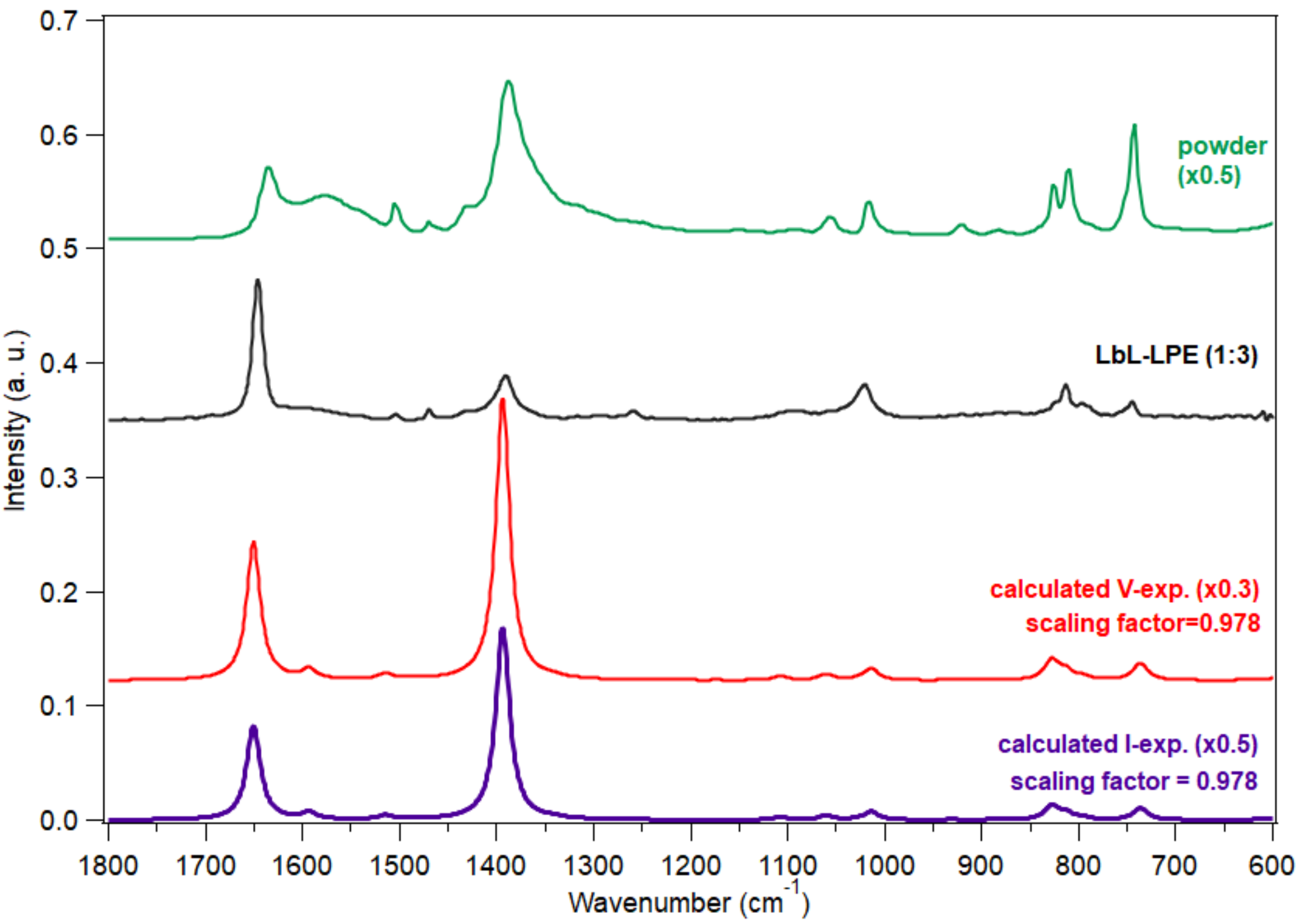

**Figure S11.** Comparison of calculated structures **IV** and **V-exp.** with the experimental spectra of the LbL-LPE film (optimal conditions) and the bulk powder. The calculated spectrum matches the bulk powder spectrum, while the oriented structure has a different intensity distribution between asymmetric and symmetric carboxylic group vibrations. This is known to occur for oriented structures, as shown for SSH-fabricated $Cu_2BDC_2DABCO$ films by Baumgartner *et al*.[24] The calculated IR spectrum was scaled (≈0.9) along the frequency axis to correct for the systematic overestimation of vibrational frequencies inherent to the harmonic approximation used in theoretical calculations, enabling better agreement with the experimental spectrum.[25]

**Table S3.** Assignment of the vibrational bands in the calculated and experimental spectra.

| **Frequency in experimental spectrum, $cm^{-1}$** | **Frequency in calculated spectrum IV, $cm^{-1}$** | **Frequency in calculated spectrum V-exp., $cm^{-1}$** | **Vibrations** |
|---|---|---|---|
| 1646 | 1610 | 1615 | $COO^-$ stretching asymmetric |
| 1588 | 1555 | 1559 | BDC stretching asymmetric |
| 1479 | 1478 | 1480 | DABCO $CH_2$ bending |
| 1391 | 1358 | 1363 | BDC stretching symmetric |
| | | | $COO^-$ stretching symmetric |
| | | | DABCO $NC_3$ stretching symmetric $CH_2$ bending |
| 1092 | 1083 | 1085 | BDC C-H scissoring |
| 1021 | 1038 | 1037 | DABCO C-C stretching asymmetric |
| 1022 | 995 | 992 | DABCO $NC_3$ scissoring |
| 812 | 810 | 810 | DABCO C-C stretching symmetric |
| 744 | 721 | 718 | BDC C-H wagging |

## 6. GIWAXS Characterization of $Zn_2(BDC)_2(DABCO)$ Films

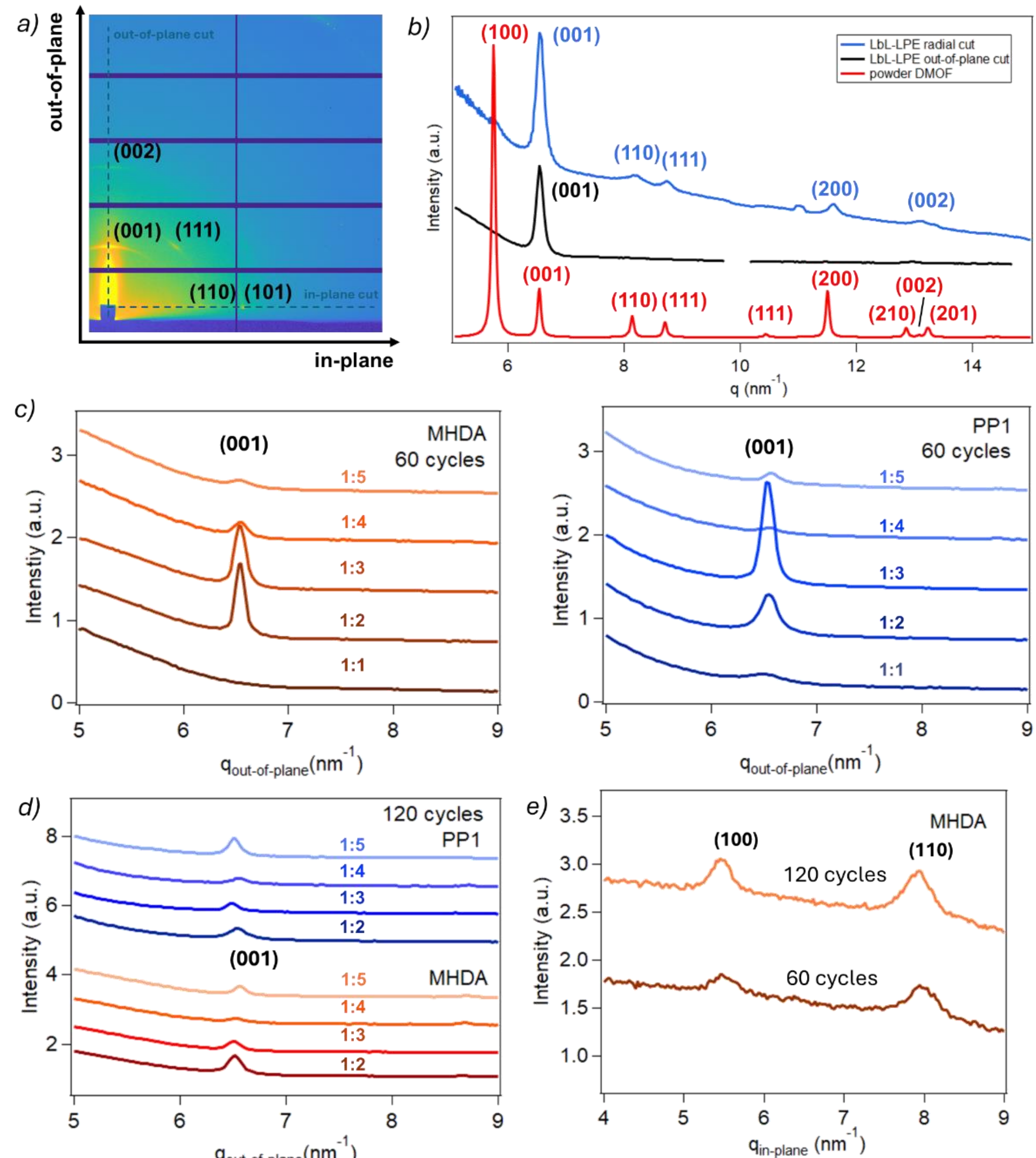

**Figure S12. Crystalline phase identification and homogeneity of the films.** (a) The LbL-LPE films show preferential orientation in the out-of-plane and in-plane direction denoted by the black arrows. The cuts employed for data analysis for the out-of-plane (cut margin 10 pixels,) and in-plane (cut margin 10 pixels) over the entire *q*-range ($q < 29.21$ nm$^{-1}$) are indicated by the black dotted line. (b) Comparison with the powder diffraction pattern of the sample prepared under optimal conditions confirms that the dominant reflection corresponds to the (001)-oriented MOF phase. The presence of the (110), (111), and (101) reflections is related to the existence of tilted crystallites. (c) Out-of-plane integration of the GIWAXS intensity of MOF thin films of 60 synthesis cycles with different molar ratios on the MHDA (left) and PP1 (right) functionalized SAM layers. These data confirm that for the molar ratio BDC:DABCO = 1:3 most of the crystallites are oriented along (001). (d) Out-of-plane integration of the GIWAXS patterns for samples with 120 fabrication cycles shows the (001) reflection, indicating that increasing the number of cycles up to 120 suppresses the crystalline orientation. (e) In-plane integration of the GIWAXS patterns of samples with the optimum ratio Zn-BDC-DABCO=1:1:3 of different synthesis cycles on MHDA layer, shows absence of a (001) in-plane alignment, which confirms that that the film is textured rather than epitaxially ordered.

## 7. Surface and cross-section morphology

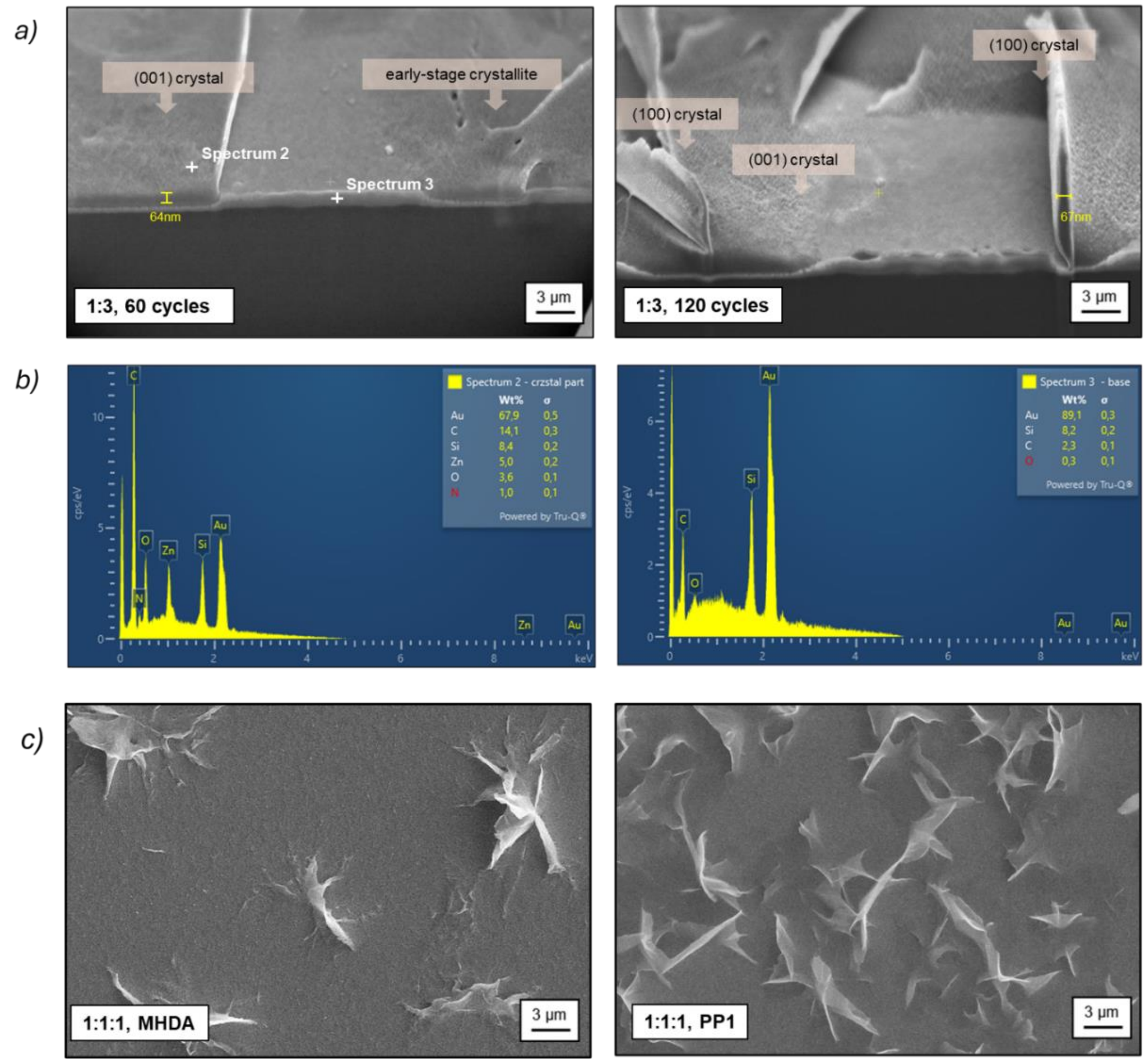

**Figure S13. SEM and EDS Analysis of MOF Film Morphology.** (a) SEM images of FIB-cut samples of 60 and 120 synthesis cycles with the molar ratio Zn:BDC:DABCO = 1:1:3 on MHDA self-assembled monolayer. (b) EDS spectra collected from two regions of the film surface in (a): on a crystal (inset, Spectrum 2) and in the area between crystals (inset, Spectrum 3). The EDS data indicate that the regions between the crystals contain no metal, suggesting that MOF growth follows an island-like mechanism. (c) SEM images of the amorphous samples with the molar ratio Zn:BDC:DABCO=1:1:1.

## 8. Thin film thickness measurement with 3D profilometry

The sputter crater depth was determined by 3D stylus profilometry from the height difference between the sputtered region and the surrounding intact surface. For each crater, the central area away from the edges was selected as the sputtered plane. These regions are marked as X1 and X2 for the sample with BDC:DABCO = 1:3 and as X3 and X4 for the sample with BDC:DABCO = 1:5. The non-sputtered surface was defined using rectangular reference areas (Y) on the surrounding intact surface. The average height of the Y regions established the reference plane, while the corresponding X region represented the crater floor. The resulting crater depths correspond to the material thickness removed during GCIB sputtering under the applied ToF-SIMS depth-profiling conditions.

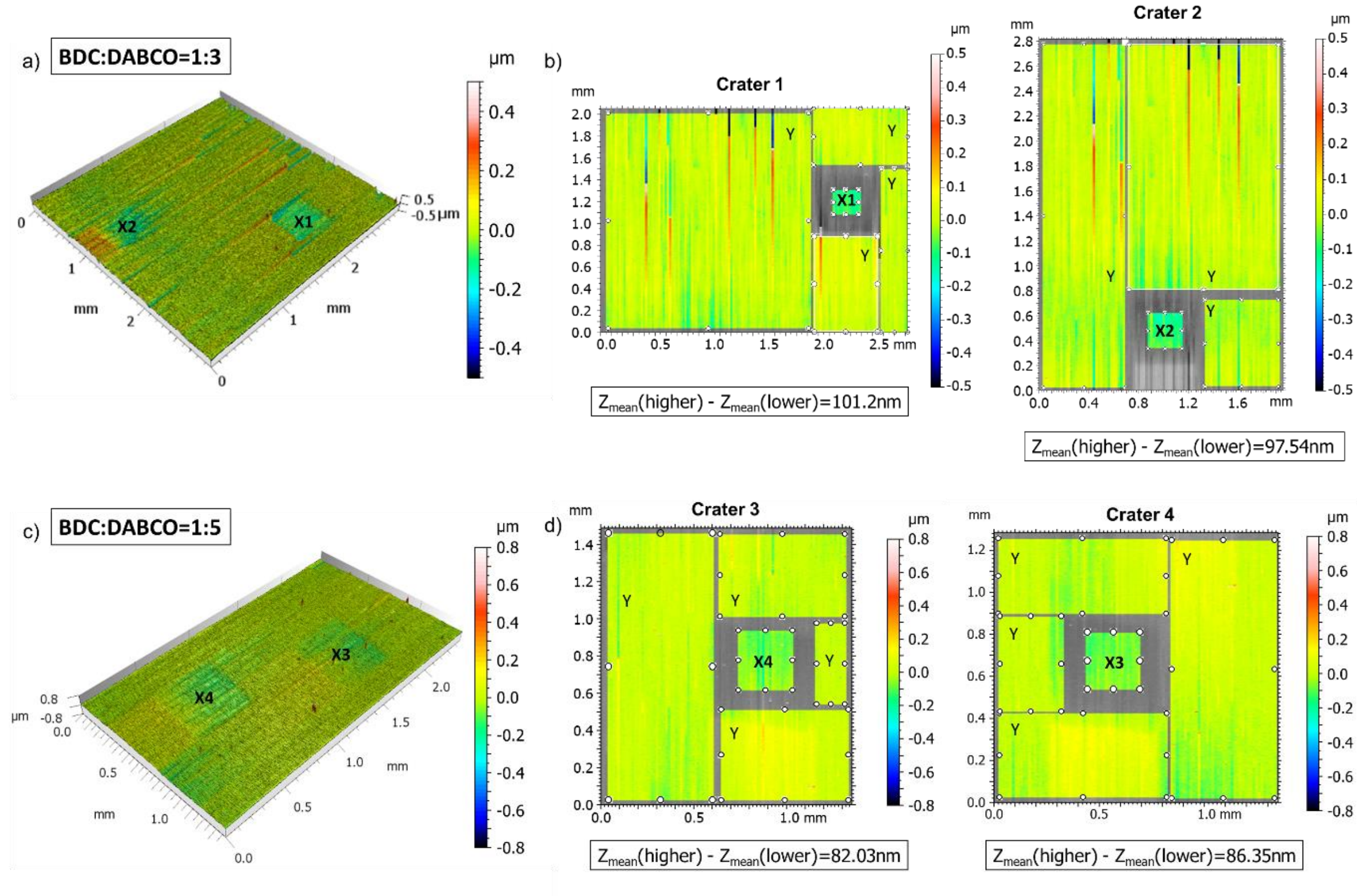

**Figure S14. Sample Thickness Determination.** The thickness of samples with 1:3 and 1:5 BDC:DABCO ligand ratios on the MHDA layer was determined by creating two craters on each sample using ToF-SIMS with GCIB sputtering, followed by measuring the height differences across the surface using 3D profilometry. Three-dimensional surface topography maps (a), (c) indicate the crater positions, while the corresponding selected surface height profiles (b), (d) are shown on the right and were used to determine the averaged crater depth extracted from the topographic data.

## 9. Degree of orientation and Hermans orientation parameter calculations

The degree of orientation (DO) and the Hermans orientation parameter (HOP) were determined by analyzing the azimuthal intensity distribution of the sharp, well-defined (001) diffraction spot, following established analysis protocols.[26–29] For a simple estimation, the flat component of the azimuthally integrated intensity was attributed to an isotropic distribution of crystallites, while the remaining intensity was assigned to preferentially oriented crystallites. This ordering metric does not account for the angular width of the preferential orientations, which can be evaluated from the azimuthal full width at half maximum (FWHM) of the respective peak.

Experimental data were reshaped into 2D diffractograms in reciprocal space using the open-access software GIXSGUI (Figure S15a).[30] To remove non-sample-related scattering, background contributions from air, the gold substrate, and other sources were estimated by taking azimuthal cuts above and below the peak, averaging them, and subtracting the result from the central cut. For quantitative texture analysis, two corrections were applied to the pole figure data (Figure S15b). First, the signal near 90° (the horizon) was excluded, as it was identified as the Yoneda peak.[31] Second, the missing wedges at χ = 0° and 37–54° were accounted for, by fitting using a Gaussian and linear functions, and the full angular range from 0° to 90° was analyzed (Figure S15c).

Correction of the pole figure was performed by multiplying the measured intensity by sin(χ). This correction converts the measured scattering intensity around $q_z$ into the total scattering signal that would be expected if all crystallites in the 2D and 3D powder textures were accessible at a given χ. Subsequent normalization of the intensity per area under the diffraction peak enables direct comparison between measurements independent of geometric factors.

The degree of orientation is calculated by

$$DO\ (\%) = \frac{A_{total} - A_{isotropic}}{A_{total}} \cdot 100$$

where $A_{total}$ is the area below the integrated intensity as a function of azimuthal angle $\chi$, and $A_{isotropic}$ is the isotropic fraction as shown in Figure S15d.[26,27]

Hermans orientation parameter (HOP) is calculated by the formulas

$$\langle cos^2\chi \rangle = \frac{\sum I(\chi) cos^2\chi}{\sum I(\chi)}$$

$$HOP = \frac{3\langle cos^2\chi \rangle - 1}{2}$$

where $I(\chi)$ is intensity at azimuthal angle $\chi$.[32]

HOP ranges from -0.5 (perfectly perpendicular alignment) to 1 (perfectly aligned along the reference axis) and 0 is fully isotropic distribution.

Although obtaining a complete pole figure down to γ = 0° is not possible for reflections along $q_z$ and corrections were therefore applied, the presented methodology provides a few-step approach for GIWAXS data analysis that enables reliable comparison of different datasets and assessment of synthesis reproducibility.

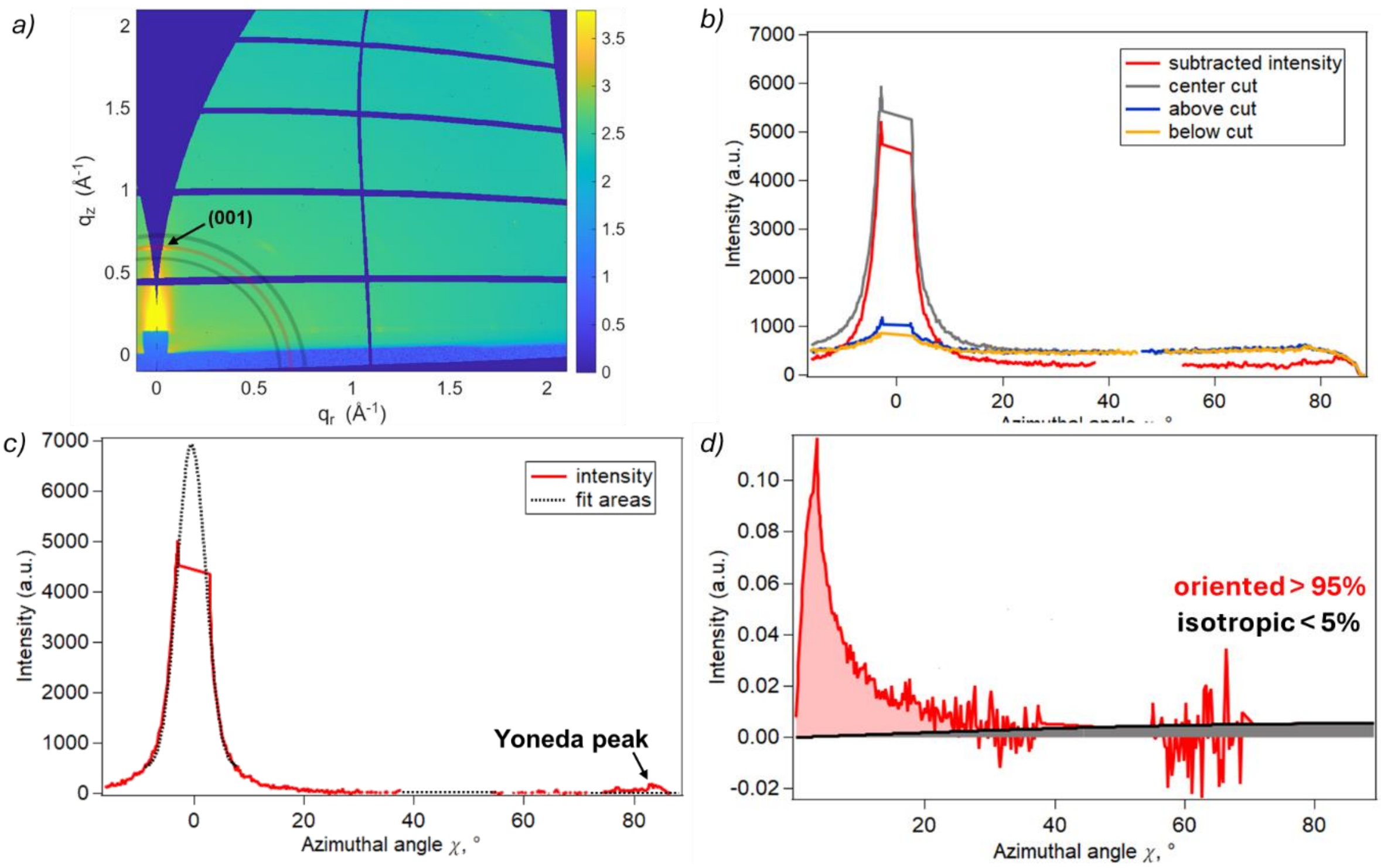

**Figure S15. GIWAXS data processing.** (a) Reshaped GIWAXS image showing the intense (001) reflection and corresponding azimuthal cuts of the background above and below the reflection, as well as the main diffraction peak. (b) Background intensity subtraction. (c) Corrections applied to the pole figure data, including missing wedge accounting and geometric weighting. (d) Corrected and area-normalized pole figure intensities highlighting the isotropic crystallite contribution.

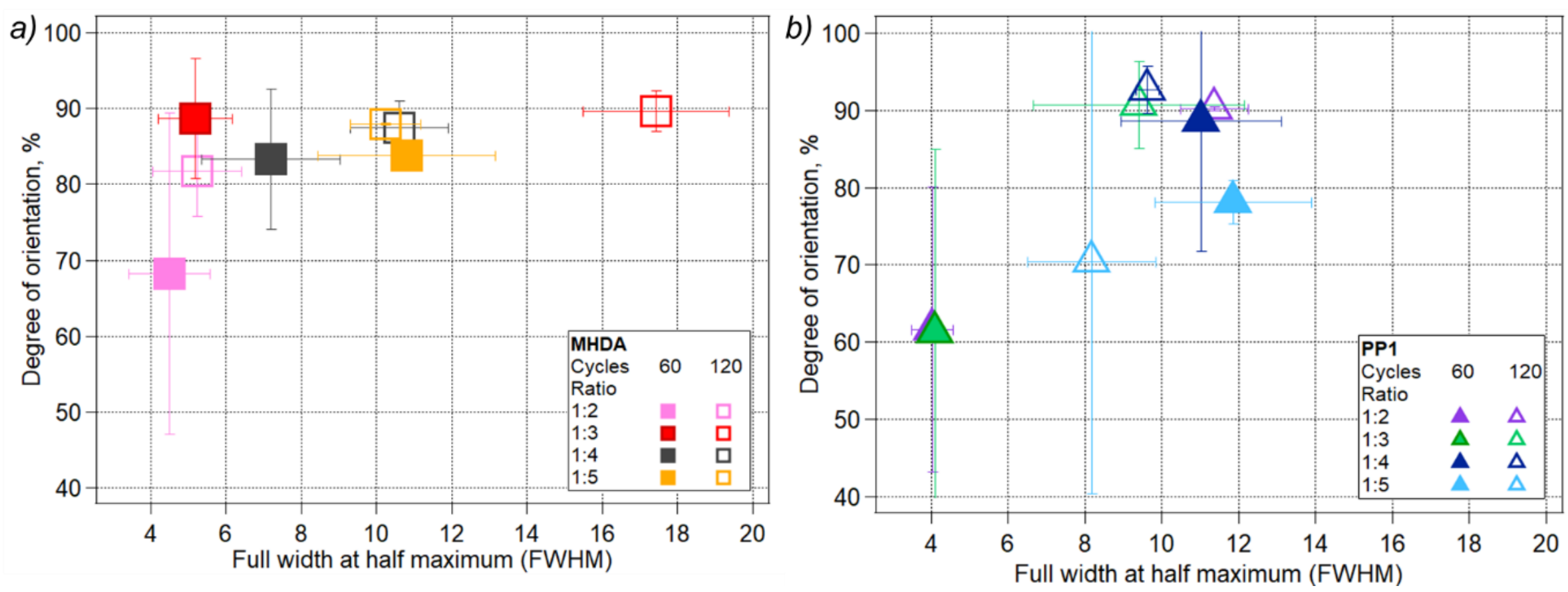

**Figure S16.** Degree of orientation (DO) versus full width at half maximum (FWHM) for LbL-LPE films grown on MHDA (a) and PP1 (b). The plot illustrates the highest degree of orientation (DO > 80%) for samples with BDC:DABCO molar ratios of 1:2 and 1:3 after 60 deposition cycles, consistent with the observed SEM data (see Figure 2 in the main manuscript).

**Table 4.** Overview of the mean values of the DO, HOP and FWHM for samples with different linker-to-linker ratio.

| Ratio | SAM | Cycles | DO | HOP | FWHM | No. of samples |
|---|---|---|---|---|---|---|
| 1:2 | MHDA | 60 cycles | 68.18 ± 21.19% | 0.86 ± 0.14 | 4.49 ± 1.08 | 3 |
| | | 120 cycles | 81.69 ± 5.94% | 0.85 ± 0.03 | 5.23 ± 1.19 | 2 |
| | PP1 | 60 cycles | 61.58 ± 18.42% | 0.94 ± 0.02 | 4.04 ± 0.54 | 3 |
| | | 120 cycles | 90.27 ± 0.28% | 0.86 ± 0.01 | 11.36 ± 0.87 | 2 |
| 1:3 | MHDA | 60 cycles | 88.64 ± 7.94% | 0.93 ± 0.04 | 5.18 ± 0.98 | 6 |
| | | 120 cycles | 89.64 ± 2.64% | 0.83 ± 0.03 | 17.42 ± 1.93 | 3 |
| | PP1 | 60 cycles | 61.30 ± 23.70% | 0.95 ± 0.03 | 4.05 ± 0.27 | 5 |
| | | 120 cycles | 90.77 ± 5.59% | 0.88 ± 0.03 | 9.39 ± 2.75 | 3 |
| 1:4 | MHDA | 60 cycles | 83.33 ± 9.23% | 0.87 ± 0.07 | 7.20 ± 1.83 | 3 |
| | | 120 cycles | 87.51 ± 3.50% | 0.78 ± 0.01 | 10.61 ± 1.31 | 2 |
| | PP1 | 60 cycles | 88.69 ± 16.91% | 0.82 ± 0.13 | 11.02 ± 2.08 | 3 |
| | | 120 cycles | 92.63 ± 3.06% | 0.84 ± 0.01 | 9.62 ± 0.28 | 2 |
| 1:5 | MHDA | 60 cycles | 83.77 ± 0.20% | 0.81 ± 0.01 | 10.79 ± 2.36 | 2 |
| | | 120 cycles | 87.91 ± 0.07% | 0.80 ± 0.01 | 10.24 ± 0.92 | 2 |
| | PP1 | 60 cycles | 78.19 ± 2.79% | 0.79 ± 0.01 | 8.18 ± 1.67 | 2 |
| | | 120 cycles | 70.40 ± 32.84% | 0.81 ± 0.02 | 9.62 ± 0.28 | 2 |

## References

(1) Wannapaiboon, S.; Schneemann, A.; Hante, I.; Tu, M.; Epp, K.; Semrau, A. L.; Sternemann, C.; Paulus, M.; Baxter, S. J.; Kieslich, G.; Fischer, R. A. Control of Structural Flexibility of Layered-Pillared Metal-Organic Frameworks Anchored at Surfaces. *Nat Commun* **2019**, *10* (1), 346. https://doi.org/10.1038/s41467-018-08285-5.

(2) Burian, M.; Meisenbichler, C.; Naumenko, D.; Amenitsch, H. *SAXSDOG* : Open Software for Real-Time Azimuthal Integration of 2D Scattering Images. *J Appl Crystallogr* **2022**, *55* (3), 677–685. https://doi.org/10.1107/S1600576722003685.

(3) Dybtsev, D. N.; Chun, H.; Kim, K. Rigid and Flexible: A Highly Porous Metal–Organic Framework with Unusual Guest-Dependent Dynamic Behavior. *Angew Chem Int Ed* **2004**, *43* (38), 5033–5036. https://doi.org/10.1002/anie.200460712.

(4) Singh, M.; Kaur, N.; Comini, E. The Role of Self-Assembled Monolayers in Electronic Devices. *J. Mater. Chem. C* **2020**, *8* (12), 3938–3955. https://doi.org/10.1039/D0TC00388C.

(5) Griffiths, K.; Halcovitch, N. R.; Griffin, J. M. Long-Term Solar Energy Storage under Ambient Conditions in a MOF-Based Solid–Solid Phase-Change Material. *Chem. Mater.* **2020**, *32* (23), 9925–9936. https://doi.org/10.1021/acs.chemmater.0c02708.

(6) Hjorth Larsen, A.; Jørgen Mortensen, J.; Blomqvist, J.; Castelli, I. E.; Christensen, R.; Dułak, M.; Friis, J.; Groves, M. N.; Hammer, B.; Hargus, C.; Hermes, E. D.; Jennings, P. C.; Bjerre Jensen, P.; Kermode, J.; Kitchin, J. R.; Leonhard Kolsbjerg, E.; Kubal, J.; Kaasbjerg, K.; Lysgaard, S.; Bergmann Maronsson, J.; Maxson, T.; Olsen, T.; Pastewka, L.; Peterson, A.; Rostgaard, C.; Schiøtz, J.; Schütt, O.; Strange, M.; Thygesen, K. S.; Vegge, T.; Vilhelmsen, L.; Walter, M.; Zeng, Z.; Jacobsen, K. W. The Atomic Simulation Environment—a Python Library for Working with Atoms. *J. Phys.: Condens. Matter* **2017**, *29* (27), 273002. https://doi.org/10.1088/1361-648X/aa680e.

(7) Blum, V.; Gehrke, R.; Hanke, F.; Havu, P.; Havu, V.; Ren, X.; Reuter, K.; Scheffler, M. Ab Initio Molecular Simulations with Numeric Atom-Centered Orbitals. *Computer Physics Communications* **2009**, *180* (11), 2175–2196. https://doi.org/10.1016/j.cpc.2009.06.022.

(8) Knuth, F.; Carbogno, C.; Atalla, V.; Blum, V.; Scheffler, M. All-Electron Formalism for Total Energy Strain Derivatives and Stress Tensor Components for Numeric Atom-Centered Orbitals. *Computer Physics Communications* **2015**, *190*, 33–50. https://doi.org/10.1016/j.cpc.2015.01.003.

(9) Ren, X.; Rinke, P.; Blum, V.; Wieferink, J.; Tkatchenko, A.; Sanfilippo, A.; Reuter, K.; Scheffler, M. Resolution-of-Identity Approach to Hartree–Fock, Hybrid Density Functionals, RPA, MP2 and *GW* with Numeric Atom-Centered Orbital Basis Functions. *New J. Phys.* **2012**, *14* (5), 053020. https://doi.org/10.1088/1367-2630/14/5/053020.

(10) Yu, V. W.; Corsetti, F.; García, A.; Huhn, W. P.; Jacquelin, M.; Jia, W.; Lange, B.; Lin, L.; Lu, J.; Mi, W.; Seifitokaldani, A.; Vázquez-Mayagoitia, Á.; Yang, C.; Yang, H.; Blum, V. ELSI: A Unified Software Interface for Kohn–Sham Electronic Structure Solvers. *Computer Physics Communications* **2018**, *222*, 267–285. https://doi.org/10.1016/j.cpc.2017.09.007.

(11) Havu, V.; Blum, V.; Havu, P.; Scheffler, M. Efficient Integration for All-Electron Electronic Structure Calculation Using Numeric Basis Functions. *Journal of Computational Physics* **2009**, *228* (22), 8367–8379. https://doi.org/10.1016/j.jcp.2009.08.008.

(12) Ihrig, A. C.; Wieferink, J.; Zhang, I. Y.; Ropo, M.; Ren, X.; Rinke, P.; Scheffler, M.; Blum, V. Accurate Localized Resolution of Identity Approach for Linear-Scaling Hybrid Density Functionals and for Many-Body Perturbation Theory. *New J. Phys.* **2015**, *17* (9), 093020. https://doi.org/10.1088/1367-2630/17/9/093020.

(13) Perdew, J. P.; Burke, K.; Ernzerhof, M. Generalized Gradient Approximation Made Simple. *Phys. Rev. Lett.* **1996**, *77* (18), 3865–3868. https://doi.org/10.1103/PhysRevLett.77.3865.

(14) Tkatchenko, A.; DiStasio, R. A.; Car, R.; Scheffler, M. Accurate and Efficient Method for Many-Body van Der Waals Interactions. *Phys. Rev. Lett.* **2012**, *108* (23), 236402. https://doi.org/10.1103/PhysRevLett.108.236402.

(15) Ambrosetti, A.; Reilly, A. M.; DiStasio, R. A.; Tkatchenko, A. Long-Range Correlation Energy Calculated from Coupled Atomic Response Functions. *The Journal of Chemical Physics* **2014**, *140* (18), 18A508. https://doi.org/10.1063/1.4865104.

(16) Togo, A. First-Principles Phonon Calculations with Phonopy and Phono3py. *J. Phys. Soc. Jpn.* **2023**, *92* (1), 012001. https://doi.org/10.7566/JPSJ.92.012001.

(17) Togo, A.; Chaput, L.; Tadano, T.; Tanaka, I. Implementation Strategies in Phonopy and Phono3py. *J. Phys.: Condens. Matter* **2023**, *35* (35), 353001. https://doi.org/10.1088/1361-648X/acd831.

(18) Skelton, J. M.; Burton, L. A.; Jackson, A. J.; Oba, F.; Parker, S. C.; Walsh, A. Lattice Dynamics of the Tin Sulphides $SnS_2$ , SnS and $Sn_2S_3$ : Vibrational Spectra and Thermal Transport. *Phys. Chem. Chem. Phys.* **2017**, *19* (19), 12452–12465. https://doi.org/10.1039/C7CP01680H.

(19) Skelton-Group/Phonopy-Spectroscopy, 2026. https://github.com/skelton-group/Phonopy-Spectroscopy (accessed 2026-01-16).

(20) *Jmol: an open-source Java viewer for chemical structures in 3D*. https://jmol.sourceforge.net/ (accessed 2026-01-16).

(21) Tang, Y.; Dubbeldam, D.; Tanase, S. Water–Ethanol and Methanol–Ethanol Separations Using in Situ Confined Polymer Chains in a Metal–Organic Framework. *ACS Appl. Mater. Interfaces* **2019**, *11* (44), 41383–41393. https://doi.org/10.1021/acsami.9b14367.

(22) Kovalenko, V. I.; Akhmadiyarov, A. A.; Vandyukov, A. E.; Khamatgalimov, A. R. Experimental Vibrational Spectra and Computational Study of 1,4-Diazabicyclo[2.2.2]Octane. *Journal of Molecular Structure* **2012**, *1028*, 134–140. https://doi.org/10.1016/j.molstruc.2012.06.045.

(23) Lugier, O.; Pokharel, U.; Castellanos, S. Impact of Synthetic Conditions on the Morphology and Crystallinity of FDMOF-1(Cu) Thin Films. *Crystal Growth & Design* **2020**, *20* (8), 5302–5309. https://doi.org/10.1021/acs.cgd.0c00529.

(24) Baumgartner, B.; Ikigaki, K.; Okada, K.; Takahashi, M. Infrared Crystallography for Framework and Linker Orientation in Metal–Organic Framework Films. *Chem. Sci.* **2021**, *12* (27), 9298–9308. https://doi.org/10.1039/D1SC02370E.

(25) Henschel, H.; Andersson, A. T.; Jespers, W.; Mehdi Ghahremanpour, M.; Van Der Spoel, D. Theoretical Infrared Spectra: Quantitative Similarity Measures and Force Fields. *J. Chem. Theory Comput.* **2020**, *16* (5), 3307–3315. https://doi.org/10.1021/acs.jctc.0c00126.

(26) Fischer, J. C.; Li, C.; Hamer, S.; Heinke, L.; Herges, R.; Richards, B. S.; Howard, I. A. GIWAXS Characterization of Metal–Organic Framework Thin Films and Heterostructures: Quantifying Structure and Orientation. *Adv Materials Inter* **2023**, *10* (11), 2202259. https://doi.org/10.1002/admi.202202259.

(27) Oesinghaus, L.; Schlipf, J.; Giesbrecht, N.; Song, L.; Hu, Y.; Bein, T.; Docampo, P.; Müller-Buschbaum, P. Toward Tailored Film Morphologies: The Origin of Crystal Orientation in Hybrid Perovskite Thin Films. *Adv Materials Inter* **2016**, *3* (19), 1600403. https://doi.org/10.1002/admi.201600403.

(28) Gasser, F.; Simbrunner, J.; Huck, M.; Moser, A.; Steinrück, H.-G.; Resel, R. Intensity Corrections for Grazing-Incidence X-Ray Diffraction of Thin Films Using Static Area Detectors. *J Appl Crystallogr* **2025**, *58* (1), 96–106. https://doi.org/10.1107/S1600576724010628.

(29) Klokic, S.; Naumenko, D.; Marmiroli, B.; Carraro, F.; Linares-Moreau, M.; Zilio, S. D.; Birarda, G.; Kargl, R.; Falcaro, P.; Amenitsch, H. Unraveling the Timescale of the Structural Photo-Response within Oriented Metal–Organic Framework Films. *Chem. Sci.* **2022**, *13* (40), 11869–11877. https://doi.org/10.1039/D2SC02405E.

(30) Jiang, Z. *GIXSGUI* : A MATLAB Toolbox for Grazing-Incidence X-Ray Scattering Data Visualization and Reduction, and Indexing of Buried Three-Dimensional Periodic Nanostructured Films. *J Appl Crystallogr* **2015**, *48* (3), 917–926. https://doi.org/10.1107/S1600576715004434.

(31) Yoneda, Y. Anomalous Surface Reflection of X Rays. *Phys. Rev.* **1963**, *131* (5), 2010–2013. https://doi.org/10.1103/PhysRev.131.2010.
(32) Lafrance, C. P.; Pezolet, M.; Prud'homme, R. E. Study of the Distribution of Molecular Orientation in Highly Oriented Polyethylene by X-Ray Diffraction. *Macromolecules* **1991**, *24* (17), 4948–4956. https://doi.org/10.1021/ma00017a035.